\documentclass[paper,a4paper,11pt,twoside]{JHEP3}
\usepackage{macros} 
\usepackage{amssymb,amsmath}
\usepackage{booktabs,lscape}
\usepackage{epsfig}
\makeatletter
\def\@xcmidrule{\ifx\@tempa\cmidrule\vskip-\@thisrulewidth
     \global\@lastruleclass=\@ne\else
     \ifx\@tempa\morecmidrules\vskip \cmidrulesep
     \global\@lastruleclass=\@ne\else
     \vskip \belowrulesep\global\@lastruleclass=\z@\fi\fi
     \ifnum0=`{\fi}}
\makeatother
\title{%
Non-perturbative renormalization\\ 
of the static axial current in two-flavour QCD
}

\author{%
\begin{flushleft}
\vspace{-0.5cm}
\vbox{
\epsfxsize=2.5 true cm
\epsfbox{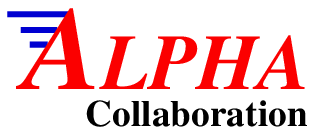}}
\end{flushleft}
}

\author{%
Michele Della Morte\\
CERN, Physics Department, TH Unit,\\ 
CH-1211 Geneva 23, Switzerland\\
E-mail: \email{dellamor@mail.cern.ch}
}
\author{%
Patrick Fritzsch, Jochen Heitger\\
Westf\"alische Wilhelms-Universit\"at M\"unster,
Institut f\"ur Theoretische Physik,\\
Wilhelm-Klemm-Strasse~9, D-48149 M\"unster, Germany\\
E-mail: \email{fritzsch@uni-muenster.de,heitger@uni-muenster.de}
}

\preprint{%
MS-TP-06-33\\
CERN-PH-TH/2006-237\\
SFB/CPP-06-52\\
\today
}

\abstract{%
We perform the non-perturbative renormalization of matrix elements of the
static-light axial current by a computation of its scale dependence in
lattice QCD with two flavours of massless O($a$) improved Wilson quarks.
The regularization independent factor that relates any running renormalized
matrix element of the axial current in the static effective theory to the 
renormalization group invariant one is evaluated in the Schr\"odinger 
functional scheme, where in this case we find a significant deviation of
the non-perturbative running from the perturbative prediction.
An important technical ingredient to improve the precision of the results
consists in the use of modified discretizations of the static quark action 
introduced earlier by our collaboration.
As an illustration how to apply the renormalization of the static axial 
current presented here, we connect the bare matrix element of the current
to the $\Bs$-meson decay constant in the static approximation for one value 
of the lattice spacing, $a\approx 0.08\,\Fm$, employing large-volume $\nf=2$ 
data at $\beta=5.3$.
}

\keywords{%
Lattice QCD, HQET, Non-perturbative renormalization, 
Improved static actions, B-meson decay constant}

\begin{document}
\section{Introduction}
\label{Sec_intro}
In view of the challenging experimental programme of B-factories and the
demand of a precise quantitative interpretation of its observations within 
or beyond the Standard Model, non-perturbative investigations of the B-meson 
system and its transition amplitudes in the framework of lattice QCD have 
become a vivid area of research.
The impact of lattice QCD on this area of flavour physics crucially depends 
on the precision that lattice computations of B-physics matrix elements are 
able to achieve.
Thus, it is very important to try to reduce its systematic errors such as 
the quenched approximation (which currently is already being overcome for 
many phenomenologically relevant 
quantities, see e.g.~\cite{lat06:kphys,lat05:ckmphys,lat06:bphys}) and the 
uncertainties owing to the still unphysically large sea quark masses 
employed in most simulations with dynamical quarks.

Yet another difficult part of these computations arises from the problem
of a sensible treatment of b-quarks on the lattice, because lattice 
spacings small enough to satisfy the condition $a<1/\mbeauty$ for a 
propagating, relativistic b-quark will certainly still continue to be out
of reach in the near future.
A theoretically clean solution is provided by the heavy quark effective 
theory (HQET). 
This is a systematic expansion of the QCD amplitudes (between hadron states
containing a single heavy quark) around the static limit, which describes 
the $\mbeauty\to\infty$ asymptotics of the effective theory in terms of 
higher-order corrections multiplied by coefficients of $\rmO(1/\mbeauty)$
and powers thereof.
For early references and more recent reviews consult 
\cite{stat:eichten,stat:eichhill1,HQET:neubert,pic03:rainer,reviews:NPRrainer_nara},
for instance. 
Among the attractive features of lattice HQET, the theoretically most 
appealing ones are \cite{lat01:mbstat,HQET:pap1}:
(i) Higher-dimensional interaction terms in the effective Lagrangian are 
treated as insertions into static correlation functions, which implies that
(ii) the continuum limit exists and results are independent of the 
regularization, and
(iii) the renormalization of the theory can be performed 
non-perturbatively, whereby also the inclusion of the
$\rmO(1/\mbeauty)$--terms along the basic strategy 
of~\cite{HQET:pap1,HQET:pap2} has recently been implemented in a concrete 
application~\cite{HQET:mb1m}.

However, even the leading (i.e.~static) approximation of HQET turns out to 
be an interesting limit, since often it is not expected to be far from 
results at the physical point and, moreover, the static results can yield 
important information for interpolating in $1/\mbeauty$ between data at 
quark masses below the b-quark mass and the static limit. 
This has been demonstrated explicitly for the case of the $\Bs$--meson 
decay constant, $\Fbs$, in quenched QCD 
in \Refs{fbstat:pap1,iechep06:BPHjochen,fbstat:pap2}, where the value of the 
decay constant in the static approximation was used to constrain the 
extrapolation of the corresponding heavy-light matrix element at finite 
quark mass values within the charm region.
With the present work we want to carry out the first step towards a removal
of the quenched approximation as one of the main systematic errors in the
aforementioned determination of $\Fbs$.
This step consists in the non-perturbative renormalization of the 
static-light axial vector current in QCD with two dynamical quark  flavours.
In contrast to lattice QCD with relativistic quarks, where the 
renormalization constant of the axial current is only a (lattice spacing
dependent) constant, the static effective theory gives rise to a scale 
dependent, multiplicative renormalization problem, which can be solved 
following the strategy of recursive finite-size scaling in an
\emph{intermediate renormalization scheme} originally proposed 
in \cite{alpha:sigma} and already employed for the corresponding quenched 
calculation in \cite{zastat:pap3}.
For a review and earlier applications of this method, we refer the reader
to \Refs{reviews:NPRrainer_nara, msbar:pap1,alpha:Nf2_2,msbar:Nf2}.

Let us briefly recall this approach to solve non-perturbative 
renormalization problems in the present context.
As for phenomenological applications one is eventually interested in matrix
elements in a scheme, in which the renormalized amplitudes of the effective
theory are matched to the QCD ones at finite quark mass, it is usually
convenient to first compute the scheme independent \emph{renormalization 
group invariant} (RGI) matrix element:
\be
\PhiRGI\equiv\ZPhi\,\ketbra{\,{\rm PS}\,}{\,\Astat\,}{\,0\,} \;.
\label{me_RGI}
\ee
Here, $\Phibare=\ketbra{{\rm PS}}{\Astat}{0}$ denotes an unrenormalized 
matrix element of the static-light axial current between a pseudoscalar 
state and the vacuum, and the renormalization factor $\ZPhi$ is such that 
it turns any bare matrix element of $\Astat$ into the RGI one.
The pseudoscalar decay constant at finite quark mass is then related to
$\PhiRGI$ through
\be
\Fps\sqrt{\mps}=
\Cps\left(M/\lMSbar\right)\times\PhiRGI\,+\,\Or\left(1/M\right) \;, 
\label{me_QCD}
\ee
where $M$ is the RGI mass of the heavy quark and $\lMSbar$ the QCD
$\Lambda$--parameter in the $\MS$ scheme.
The function $\Cps$ in \eq{me_QCD} accounts for the fact that in order to 
extract predictions for QCD from results computed in the effective theory, 
its matrix elements are to be linked to the corresponding QCD matrix 
elements at finite values of the quark mass.
In this sense $\Cps$ translates to the 
\emph{`matching scheme'} \cite{zastat:pap3}, which is \emph{defined} by the 
condition that matrix elements in the (static) effective theory, 
renormalized in this scheme and at scale $\mu=\mbeauty$, equal those in QCD 
up to $1/\mbeauty$--corrections.
Thanks to the three-loop calculation of the anomalous dimension of the 
static axial current in the $\MS$ scheme \cite{ChetGrozin}, the function 
$\Cps(M/\lMSbar)=\Fps\sqrt{\mps}/\PhiRGI$ is known perturbatively up to and 
including $\gbar^4(\mb)$--corrections.
Therefore, the remaining perturbative uncertainty induced in (\ref{me_QCD}) 
by the conversion factor $\Cps$ is already below the 2\% level beyond the 
charm threshold and thus very small.

Assuming that the running matrix element $\Phi(\mu)$ has been
non-perturbatively defined in an intermediate renormalization scheme where
$\mu=1/\lmax$ represents a low-energy scale, the total renormalization 
factor $\ZPhi$ in \eq{me_RGI} splits into a universal and regularization 
dependent factor according to
\bea
\PhiRGI
&    =   &
\left.\frac{\PhiRGI}{\Phi(\mu)}\,\right|_{\,\mu=1/\lmax}
\times\left.\zastat(g_0,L/a)\,\right|_{\,L=\lmax}\times\Phibare(g_0)
\nonumber\\[1.0ex]
& \equiv & 
\ZPhi(g_0)\,\Phibare(g_0) \;.
\label{PhiRGI_ZRGI}
\eea
The computation of $\ZPhi(g_0)$ is the main goal of this work.

The paper is organized as follows.
In \sect{Sec_rscheme} we present our intermediate renormalization scheme,
formulated in terms of the QCD Schr\"odinger functional.
\Sect{Sec_running} contains the numerical determination of the 
(lattice formulation independent) scale dependence of the current in this 
scheme, which is the key prerequisite in order to relate the current 
renormalized at some proper low scale to the RGI current, 
while \sect{Sec_match} gives our results for the (lattice formulation 
dependent) values of the $Z$--factor at this low scale. 
In \sect{Sec_MEfinite} we then explain the use of our results and, as a 
further illustration, combine them with $\nf=2$ data for the bare matrix 
element of the static axial current from ongoing large-volume 
simulations~\cite{fbstat:Nf2} to extract $\Fbsstat$ for one value of the 
lattice spacing, $a\approx 0.08\,\Fm$.
We conclude with a discussion of the results in \sect{Sec_concl}. 
Some technical details and tables with parts of the numerical results
are deferred to appendices.

\section{The renormalization scheme}
\label{Sec_rscheme}
We consider QCD with two flavours of mass-degenerate dynamical sea quarks,
and the heavy quark field is treated in leading order of HQET 
(static approximation).
The renormalization pattern of an arbitrary matrix element $\Phibare$ of 
the heavy-light axial vector current,
\be
\Astat(x)=\lightb(x)\gamma_0\gamma_5\heavy(x) \;,
\label{astat}
\ee
is characterized by the fact that --- owing to the absence of the axial 
Ward identity which holds for the corresponding relativistic current --- 
the static-light axial current picks up a scale dependence upon 
renormalization.
Consequently, the scale evolution of the renormalized matrix element
\be
\Phi(\mu)\equiv
\ketbra{\,{\rm PS}\,}{\,\Aren(\mu)\,}{\,0\,}=
\zastat(\mu)\ketbra{\,{\rm PS}\,}{\,\lightb\gamma_0\gamma_5\heavy\,}{\,0\,}
\label{me_Phi}
\ee
between a static-light pseudoscalar state and the vacuum is governed by the 
renormalization group equation
\be
\mu\,{{\rmd\Phi}\over{\rmd\mu}}=\gamma(\gbar)\Phi
\label{rge_Phi}
\ee
in formally the same way as it is encountered in conjunction with the
running of the renormalized quark masses in QCD.

In the simple form of the renormalization group equation (\ref{rge_Phi})
we have already implicitly assumed that a mass-independent renormalization 
scheme is chosen, which is equivalent to the prescription of imposing
renormalization conditions at zero quark mass \cite{Weinberg:1951ss}.
Moreover, when introducing the lattice spacing $a$ as the regulator of the
theory, the renormalization factor in question becomes a function of the
bare coupling $g_0$ and $a\mu$, $\zastat=\zastat(g_0,a\mu)$, and only in
renormalized quantities this regulator can be removed by taking the 
continuum limit $a\to0$ to finally obtain finite results.
In the same way as for the renormalized quark masses, also in the static 
effective theory considered here the crucial advantage of mass-independent 
renormalization schemes is that in all such schemes the ratios of 
renormalized matrix elements constructed as in \eq{me_Phi} but with a 
different flavour content are scale and scheme independent constants.

The anomalous dimension associated with the renormalization scale 
dependence of the static-light axial current is encoded in the
renormalization group function $\gamma$ appearing in \eq{rge_Phi}, the
perturbative expansion of which reads
\be
\gamma(\gbar)
\,\,_{\mbox{
$\stackrel{\displaystyle\sim}{\scriptstyle\gbar\rightarrow0}$}}\,
-\gbsq\,\Big\{\,
\gamma_0+\gamma_1\gbsq+\gamma_2\gbar^4+\Or(\gbar^6)\,\Big\} \;,
\label{gam_pert}
\ee
with a universal, scheme independent 
coefficient \cite{Shifman:1987sm,Politzer:1988wp}
\be
\gamma_0= 
-{1\over{4\pi^2}}
\label{gam_0}
\ee
and higher-order ones $\gamma_1,\gamma_2,\ldots$ depending on the chosen 
renormalization scheme.
Note, however, that generically the $\gamma$--function is 
non-perturbatively defined as long as this is the case also for the matrix
element $\Phi$ of the current as well as for the renormalized gauge
coupling $\gbar$ itself.

As already emphasized in \sect{Sec_intro}, we advocate a strategy that
regards the \emph{renormalization group invariants} (RGIs) as the essential
physical objects of interest, because these are the quantities whose total 
dependence on the renormalization scale $\mu$ vanishes.
For the present investigation, the relevant RGIs are the renormalization 
group invariant counterpart (\ref{me_RGI}) of the matrix 
element (\ref{me_Phi}) of the static-light axial current, 
\be
\PhiRGI= 
\Phi(\mu)
\left[\,2b_0\gbar^2\,\right]^{-\gamma_0/(2b_0)}
\exp\left\{-\int_0^{\gbar} \rmd g 
\left[\,{\gamma(g)\over\beta(g)}-{\gamma_0 \over b_0 g}\,\right]
\right\} \;,
\label{me_PhiRGI}
\ee
and the QCD $\Lambda$--parameter
\be
\Lambda=
\mu\left[\,b_0\gbar^2\,\right]^{-b_1/(2b_0^2)}
\Exp^{\,-1/(2b_0\gbar^2)}
\exp\left\{-\int_0^{\gbar} \rmd g
\left[\,\frac{1}{\beta(g)}+\frac{1}{b_0g^3}-\frac{b_1}{b_0^2g}\,\right]
\right\} \;,
\label{Lambda}
\ee
where the renormalized coupling $\gbar=\gbar(\mu)$ obeys
\be
\beta(\gbar)=
\mu\,{{\rmd\gbar}\over{\rmd\mu}}
\,\,_{\mbox{
$\stackrel{\displaystyle\sim}{\scriptstyle\gbar\rightarrow0}$}}\,
-\gbar^3\,\Big\{\,
b_0+b_1\gbsq+b_2\gbar^4+\Or(\gbar^6)\,\Big\}
\ee
with scheme independent one- and two-loop coefficients
\be
b_0=\frac{1}{(4\pi)^2}\left(11-\frac{2}{3}\Nf\right) \;,\quad
b_1=\frac{1}{(4\pi)^4}\left(102-\frac{38}{3}\Nf\right) \;.
\ee
Both, $\PhiRGI$ as well as $\Lambda$ are defined independent of 
perturbation theory, and particularly the former is not only scale but 
also \emph{scheme} independent.
\subsection{Static-light correlation functions in the Schr\"odinger functional}
A convenient mass-independent renormalization scheme, which has already 
proven to be a theoretically attractive as well as numerically efficient 
framework to solve renormalization problems in QCD similar to that studied
here \cite{zastat:pap3,msbar:pap1,alpha:Nf2_2,msbar:Nf2,bk:pap1}, is 
provided by the QCD Schr\"odinger functional 
(SF) \cite{SF:LNWW,SF:stefan1,SF:stefan2}.
It is defined through the partition function of QCD on a $T\times L^3$ 
cylinder in Euclidean space, $\mathcal{Z}
=\int_{T\times L^3}\rmD[U,\psibar,\psi]\,\rme^{\,-S[U,\psibar,\psi]}$, where in 
the lattice regularized form we integrate over SU(3) gauge fields $U$ with 
the Wilson action and two flavours of O($a$) improved Wilson quarks, 
$\psi,\psibar$.
At times $x_0=0,T$ Dirichlet boundary conditions are imposed on the 
gluon and quark fields, whereas in the spatial directions of length $L$ 
the fields satisfy periodic (for the quark fields only up to a global phase 
$\theta$ \cite{pert:1loop}) boundary conditions.
Particularly the Dirichlet boundary conditions in time qualify the SF as a 
mass independent renormalization scheme, since they introduce an infrared 
cutoff to the frequency spectrum of quarks and gluons and hence allow to 
perform simulations at zero quark mass.
The settings $T=L$, $\theta=0.5$ and vanishing boundary gauge fields at
$x_0=0,T$ then complete the specification of our (intermediate) 
finite-volume renormalization scheme, in which the running renormalization 
scale $\mu$ is now identified with $1/L$ in a natural way.

For the non-perturbative renormalization of the static-light axial current
in two-flavour QCD we closely follow the quenched 
calculation detailed in \cite{zastat:pap3}. 
Here, we only recall the definition of the basic correlation functions
in continuum notation,
\bea
\fastat(x_0)
& = & 
-\frac{1}{2}\intdd{3}{\vecy}\dd{3}{\vecz}
\evalbig{\Astat(x)\,\zetahb(\vecy)\gamma_5\zetal(\vecz)} \;, 
\label{fastat} \\
\fdeltaAstat(x_0) 
& = & 
-\frac{1}{2}\intdd{3}{\vecy}\dd{3}{\vecz}
\evalbig{\delta\Astat(x)\,\zetahb(\vecy)\gamma_5\zetal(\vecz)} \;,
\label{fdastat} \\
\fonestat 
& = &
-\frac{1}{2L^6}\intdd{3}{\vecu}\dd{3}{\vecv}\!\dd{3}{\vecy}\!\dd{3}{\vecz}
\evalbig{\zetalbprime(\vecu)\gamma_5\zetahprime(\vecv)\,
\zetahb(\vecy)\gamma_5\zetal(\vecz)} \;,
\label{f1hl}
\eea
in terms of the static current (\ref{astat}), its $\Or(a)$ correction
\be
\delta\Astat(x)=
\lightb(x)\gamma_j\gamma_5{1\over 2}
\left(\lnab{j}+\lnabstar{j}\right)\heavy(x)
\label{dastat}
\ee
and the `boundary quark and antiquark fields', $\zeta,\zetabar$, 
the proper definition of which can be inferred 
e.g.~from \Refs{impr:pap1,zastat:pap1}.
These correlators are schematically depicted in~\fig{fig:cyl_fXstat}.
%
\FIGURE[t]{
\epsfig{file=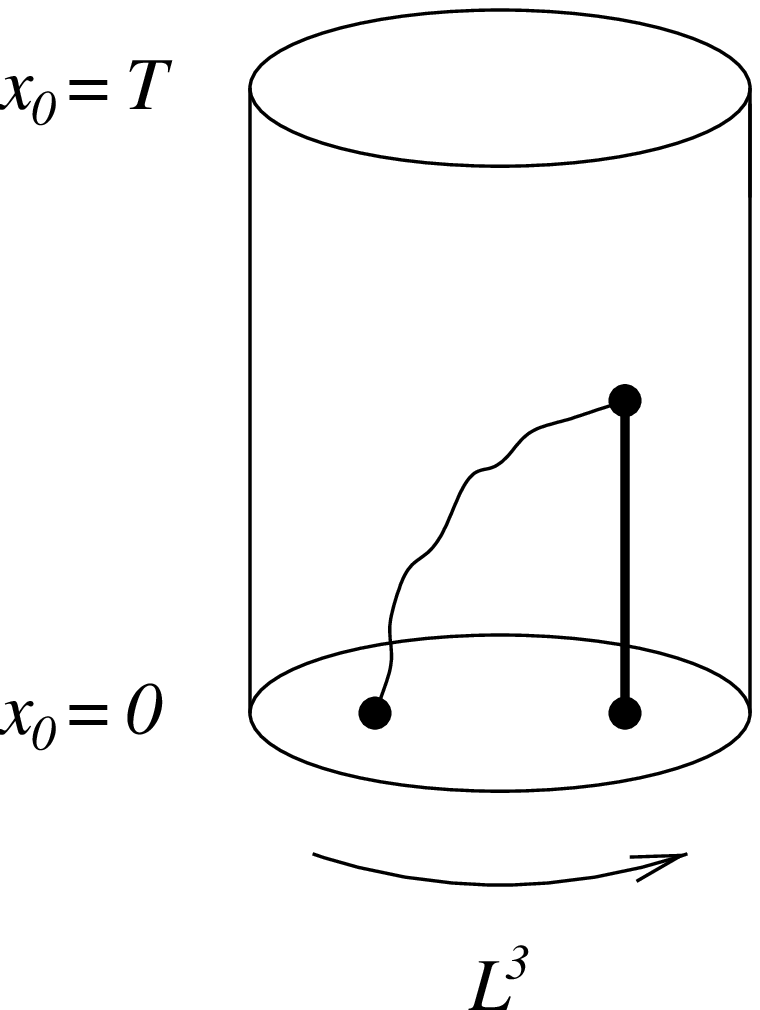,width=4.0cm}
\hspace{2.0cm}
\epsfig{file=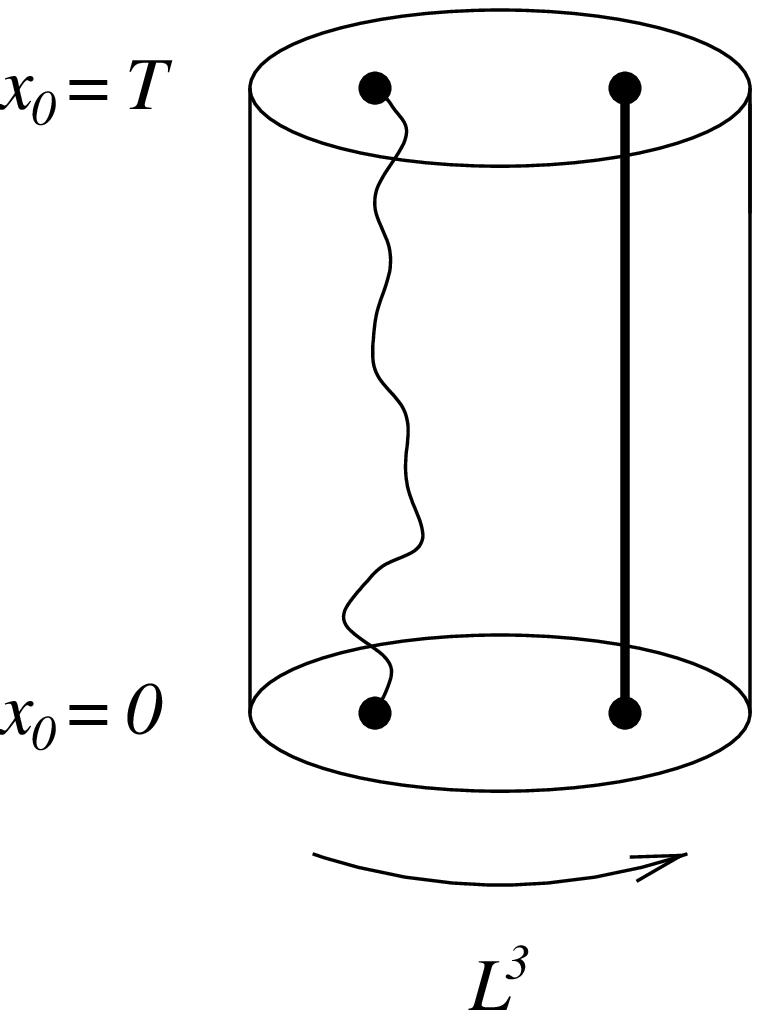,width=4.0cm}
\caption{
Illustration of the correlation functions $\fastat(x_0)$ (left) and 
$\fonestat$ (right), defined within the Schr\"odinger functional.
The curly and straight lines represent light (i.e.~relativistic) and 
static quark propagators, respectively.
In the left diagram, the static current $\Astat$ is understood to be
inserted in the bulk of the SF cylinder, at the point where the two
quark lines meet.
For $\fdeltaAstat(x_0)$, $\Astat$ is replaced by $\delta\Astat$.
}\label{fig:cyl_fXstat}
}
%
$\fastat(x_0)$, for instance, can be shown to be proportional to the matrix 
element of the (static) axial current that is inserted at time distance
$x_0$ from a pseudoscalar boundary state, while the state at the other 
boundary has vacuum quantum numbers \cite{msbar:pap2}.
For the corresponding formulae in the lattice regularized theory, in 
which eqs.~(\ref{fastat})~--~(\ref{f1hl}) and any expression derived 
therefrom receive a precise meaning, we again refer 
to \Refs{zastat:pap3,zastat:pap1}.
\subsection{Normalization condition for the static axial current}
As renormalization condition for the static axial current we impose the
condition originally formulated in the perturbative context
of~\cite{zastat:pap1} and later also explored within the non-perturbative
computation of~\cite{zastat:pap3} in the quenched approximation of QCD.
Switching to the lattice notation from now on, it reads
\be
\zastat(g_0,L/a)\,X_{\rm I}(g_0,L/a)=X_{\rm I}(0,L/a) \;,
\label{rencond}
\ee
where $X_{\rm I}(g_0,L/a)$ denotes a suitable $\Or(a)$ improved ratio 
composed from the correlation functions~(\ref{fastat})~--~(\ref{f1hl}):
\be
X_{\rm I}(g_0,L/a)\equiv
\frac{\fastat(L/2)+a\,\castat\fdeltaAstat(L/2)}{\sqrt{\fonestat}} \;.
\label{XIdef}
\ee
In this construction, the boundary-to-boundary correlator $\fonestat$
serves to cancel the unknown wave function renormalization factors of the 
boundary quark fields as well as the linearly divergent mass counterterm 
$\delta m$ that one is usually faced with in the static effective theory.
The definition of $\zastat$ via the condition (\ref{rencond}) is such that
$\zastat=1$ at tree-level of perturbation theory\footnote{
The numerical values for the tree-level normalization constant 
$X_{\rm I}(0,L/a)$ can be found in table~4 of Appendix~A in 
\Ref{zastat:pap3}.
}
(i.e.~$\zastat(0,L/a)=1$); restricting the discussion to the Eichten-Hill 
action for the static quark \cite{stat:eichhill1} for the moment, the 
improvement coefficient $\castat$ takes the one-loop perturbative 
value \cite{stat:ca,stat:ca_jp}
\be
\castat=-1.00(1)\,{1\over 4\pi}\,g_0^2 \;. 
\label{ca_stat}
\ee

Moreover, to comply with the demand of setting up a mass independent 
scheme, \eq{rencond} is to be supplemented by the condition of vanishing 
light quark mass,
\be
m_0=\mc \;,
\label{m_crit}
\ee
with bare and critical quark masses as defined in \cite{zastat:pap1}.
The critical quark mass fixes the hopping parameter, at which the 
normalization condition for the static axial current at given $L/a$ and 
$\beta=6/g_0^2$ is to be evaluated.
Its numerical values were already determined in \Ref{alpha:Nf2_2} from the 
non-perturbatively $\Oa$ improved PCAC mass in the light quark sector
(the latter being defined for $\theta=0.5$ through an appropriate 
combination of light-light correlation functions calculated at $x_0=L/2$) 
and have also been used before in the computation of the running of the 
quark mass in two-flavour QCD \cite{msbar:Nf2}.

From the normalization condition (\ref{rencond}), together with the 
one-to-one correspondence between pairs $(L/a,\beta)$, pertaining to a 
certain fixed value of the renormalized SF gauge coupling $\gbar^2(L)$, and 
the box size $L$ as the only physical scale in the system, it is obvious 
that the renormalization constant $\zastat$ runs with the scale $\mu=1/L$.
Therefore, the change of the matrix element $\Phi$ in \eq{me_Phi}, 
renormalized in the SF scheme, under finite changes of the renormalization 
scale can now be non-perturbatively computed by means of an associated 
\emph{step scaling function} $\sigmaAstat$, which is defined by the change 
induced by a scale factor of two, viz.
\be
\sigmaAstat(u)=
\frac{\Phi(\mu/2)}{\Phi(\mu)}=\frac{\zastat(2L)}{\zastat(L)} \;;
\label{SSF_cont}
\ee
in the continuum limit, it only depends on the renormalized SF coupling 
$u\equiv\gbar^2(L)$.

The (non-universal) two-loop coefficient of the anomalous dimension 
(\ref{gam_pert}) of the static axial current renormalized in the particular
SF scheme specified in this subsection is known from \Ref{zastat:pap1}
to be
\be
\gamma_1=\gamSF_1=
\frac{1}{(4\pi)^2}\,\Big(\,0.08(2)-0.0466(13)\Nf\,\Big)
\label{gam_1}
\ee
and will enter the numerical evaluation of the RGI matrix 
element (\ref{me_PhiRGI}) later.

Before we come to explain the lattice computation of $\sigmaAstat(u)$ and
the subsequent steps to relate a bare matrix element of the current to the 
RGI one, let us comment on a difference to the quenched computation 
of~\cite{zastat:pap3}.
There it turned out that in case of the usual Eichten-Hill action for the
static quark the lattice step scaling function $\SigmaAstat(u,a/L)$ 
(cf.~\eq{SSF_lat}) extracted from Monte Carlo simulations acquires large 
statistical errors in the relevant coupling range of $u\gtrsim 1.5$ and 
that these even grow drastically with $L/a$.
This fact originates from the noise-to-signal ratio of the 
boundary-to-boundary correlator $\fonestat$ in the renormalization 
condition (\ref{rencond}), which roughly behaves as 
$\,\exp\left\{e^{(1)}g_0^2\times(T/a)\right\}$ due to the self-energy of a 
static quark propagating over a distance $T=L$.
Here, $e^{(1)}$ is the leading coefficient of the linearly divergent
binding energy $E_{\rm stat}$ of the static-light system,
$E_{\rm stat}\sim\frac{1}{a}\,e^{(1)}\,g_0^2+\ldots\,$, and one infers that 
the precision problem of $\SigmaAstat$ becomes more severe towards the 
continuum limit, particularly for the Eichten-Hill action having a rather 
large coefficient, 
$e^{(1)}=\frac{1}{12\pi^2}\times19.95$ \cite{stat:eichhill_za}.
Thus, for the \emph{quenched} non-perturbative computation, the scheme 
specified above was finally discarded in favor of a slightly adapted 
scheme \cite{zastat:pap3}, in which $\fonestat$ is replaced by a product of
boundary-to-boundary correlators involving a light and a static 
quark-antiquark pair, respectively, whereby especially the latter could
be calculated with small statistical errors by applying the variance 
reduction method of~\Ref{PPR} that consists in estimating the arising 
one-link integrals by a multi-hit procedure.
In the case of QCD with \emph{dynamical} quarks, however, multi-hit can not 
be used, since it does not yield an unbiased estimator any more and 
--- being a stochastical procedure rather than analytically defined ---
it can not be traded for a change in the discretization of the action for 
the static quark.

Instead of recoursing to an alternative combination of correlators as done
in the quenched case \cite{zastat:pap3}, we therefore have to pursue a 
different direction in order to overcome the exponential degradation of the 
signal-to-noise ratio of static-light correlation functions computed with
the Eichten-Hill lattice action while maintaining the original 
renormalization condition, \eq{rencond}.
Fortunately enough, this is indeed possible thanks to the alternative 
discretizations of the static theory devised 
in~\Refs{fbstat:pap1,HQET:statprec}, which lead to a substantial gain in 
numerical precision of B-meson correlation functions in lattice HQET.

\section{The running of the renormalized static axial current}
\label{Sec_running}
As emphasized at the end of the previous section, our lattice calculations
employ alternative discretizations of the static theory that retain the 
$\Or(a)$ improvement properties of the Eichten-Hill 
action \cite{stat:eichhill1} but entail a large reduction of the statistical
fluctuations of heavy-light correlation functions with B-meson quantum 
numbers \cite{fbstat:pap1,HQET:statprec}.
In the following, we present results from the static actions denoted as 
$S^\text{s}\lh$, $S^\text{HYP1}\lh$ and $S^\text{HYP2}\lh$, or for short, `s', 
HYP1 and HYP2, respectively.
The form and a few properties of these lattice actions are briefly 
summarized in~\App{App_lattice}.

The light quark sector is represented by non-perturbatively $\Or(a)$ 
improved dynamical Wilson quarks, and we refer to \Ref{msbar:Nf2} for any 
unexplained details.
In particular, the dynamical fermion configurations, which were generated 
in the context of that reference for a series of given renormalized SF 
couplings $\gbar^2(L)$ at various lattice resolutions, constitute the basis 
for the numerical evaluation of the SF correlation functions 
(\ref{fastat})~--~(\ref{f1hl}) and the renormalization constant $\zastat$ 
via \eq{rencond}.

Note that $\zastat=\zastat(g_0,L/a)$ now carries a dependence on the type 
of static quark action used.
This holds true also for the step scaling function deduced from it, unless 
an extrapolation to the continuum limit is eventually performed 
(universality). 
\clearpage
\subsection{Continuum extrapolation of the step scaling function}
\label{Sec_running_CL}
%
\FIGURE[t]{
\epsfig{file=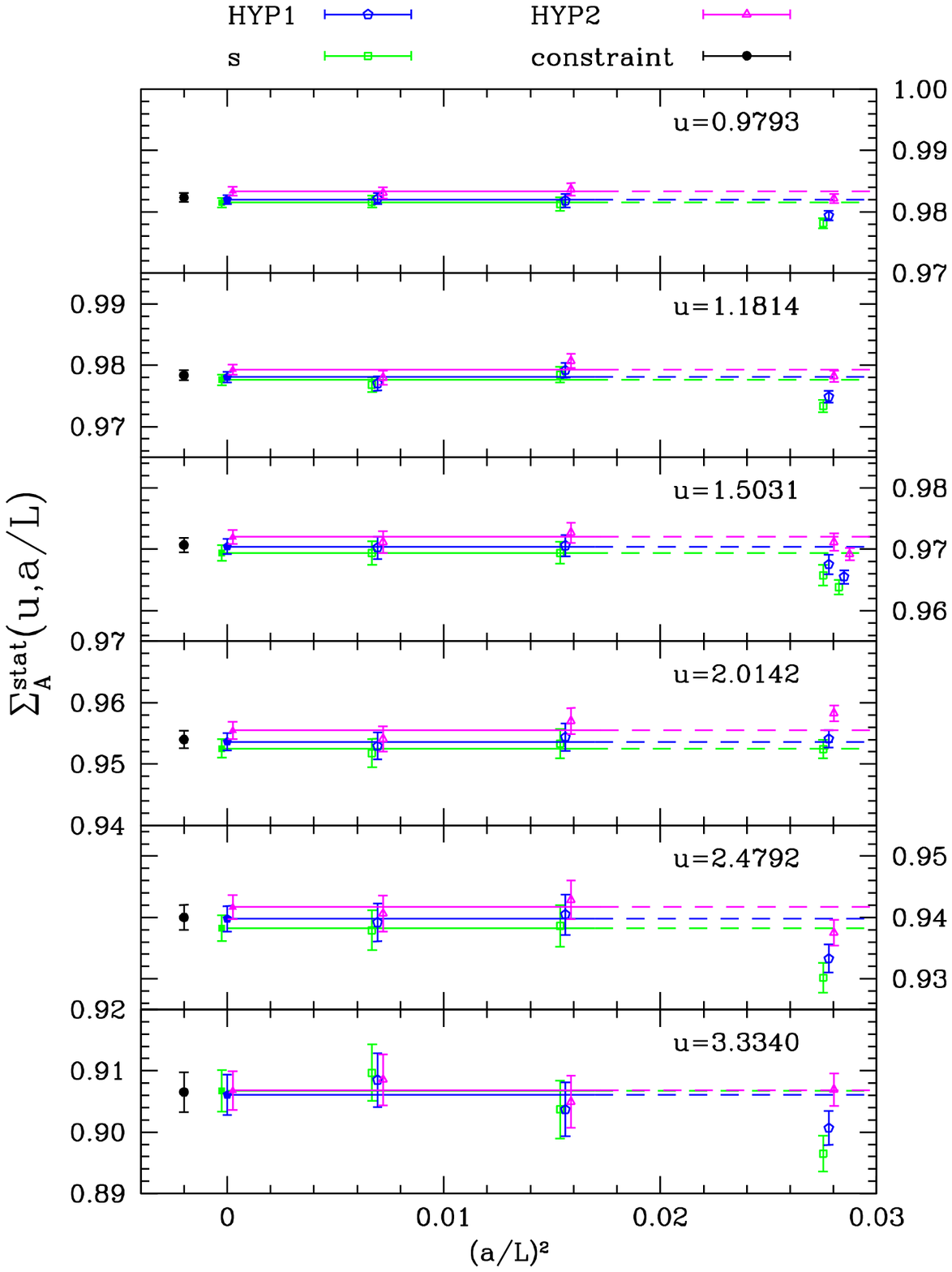,width=11.0cm}
\caption{
The lattice step scaling function $\SigmaAstat(u,a/L)$ and its continuum 
limit extrapolations to a constant omitting the $L/a=6$ data, separately 
for the three different static discretizations 
$S^\text{s}\lh$, $S^\text{HYP1}\lh$ and $S^\text{HYP2}\lh$ (colored lines).
The black points (slightly shifted to the left of $a=0$) refer to the
corresponding constrained $\chi^2$--minimization of the data from all
three actions and give our final results for $\sigmaAstat(u)$ 
in \tab{tab:sigastat_res}.
For the third smallest coupling, the two sets of points at $L/a=6$ refer 
to one-loop and two-loop values for the boundary improvement coefficient 
$\ct$ \cite{alpha:SU3,impr:ct_2loop}.
}\label{fig:Sigma_zastat}
}
%
\TABLE{
\centering
\renewcommand{\arraystretch}{1.25}
\begin{tabular}{ccl} \toprule
  $u$ & & \clmc{$\sigmaAstat(u)$} \\ \midrule 
  $0.9793$ & & $0.9823(7)(2)  $ \\
  $1.1814$ & & $0.9784(8)(11) $ \\ 
  $1.5031$ & & $0.9707(12)(3) $ \\ 
  $2.0142$ & & $0.9540(14)(9) $ \\ 
  $2.4792$ & & $0.9400(20)(4) $ \\ 
  $3.3340$ & & $0.9066(33)(25)$ \\ \bottomrule%
\end{tabular}
\caption{
Results for the continuum step scaling function $\sigmaAstat(u)$ from
constrained fits to a constant, excluding the $L/a=6$ data.
The first error is statistical, while the second one is the difference 
between the fit and the $L/a=8$ result and is accounted for as a 
systematic error.
}\label{tab:sigastat_res}
}
%
With \eq{SSF_cont} of \sect{Sec_rscheme} it was already anticipated that the
running of a renormalized matrix element of the static axial current in the
SF scheme with $\Nf=2$ massless quark flavours can be extracted from the
step scaling function $\sigmaAstat(u)$, which is defined as the continuum 
limit of the lattice step scaling function $\SigmaAstat(u,a/L)$, i.e.
\be
\sigmaAstat(u)=
\lim\limits_{a\to 0}\SigmaAstat(u,a/L) \;,\quad
\SigmaAstat(u,a/L)=
\left.{{\zastat(g_0,2L/a)}\over{\zastat(g_0,L/a)}}\,\right|_{
\,\gbar^2(L)=u\,,\,m_0=\mc} \;. 
\label{SSF_lat}
\ee
The condition $m_0=\mc$, \eq{m_crit}, refers to lattice size $L/a$ and
amounts to set the hopping parameter in the simulations to its critical
value, $\kappa=\hopc$.
Enforcing $\gbar^2(L)$ to take some prescribed value $u$ fixes the bare 
coupling value $g_0^2=6/\beta$ to be used for given $L/a$.
In this way, $\SigmaAstat$ becomes a function of the renormalized coupling
$u$ up to cutoff effects and approaches its continuum limit as 
$a/L\rightarrow 0$ for fixed $u$.

We computed $\zastat(u,a/L)$ at six values of $u$, where the corresponding
box sizes cover a range of the order $L=0.01\,\Fm\,\ldots\,1\,\Fm$ 
(or $\mu$ in $20\,\GeV\,\ldots\,0.2\,\GeV$).\footnote{
Note that the exact value of the scale in physical units does not affect 
the determination of the renormalization factors in question in this work.
As will become clear in the following subsections, it is enough to specify
the value of the SF coupling for a certain maximal box size in the hadronic 
regime.   
}
At each $u$, three lattice resolutions --- $L/a=6,8,12$ --- were simulated,
and the numerical results for $\zastat$ and $\SigmaAstat$ stemming from the
three static discretization s, HYP1 and HYP2 are collected 
in \tabs{tab:zastat1_res} and \ref{tab:zastat2_res} in \App{App_results}.
Statistical errors were estimated by a jackknife analysis and cross-checked
with the method of \Ref{MCerr:ulli}.

Given the available data sets belonging to the static actions 
$i\in\{\text{s},\text{HYP1},\text{HYP2}\}$ as well as the fact that we
work in (modulo $\castat$, see \App{App_lattice}) 
\emph{non}-pertur\-ba\-tive\-ly $\Or(a)$ improved QCD, there are basically 
two different ways to extrapolate the lattice step scaling function to the 
continuum. 
Either one can perform separate fits
\be
\Sigma^\text{stat}_{\text{A,}i}(u,a/L)= 
\sigma^\text{stat}_{\text{A,}i}(u)+\rho_i(u)\times(a/L)^2
\label{extrap_sep}
\ee
or, assuming universality of the continuum limit, a simultaneous 
extrapolation under the constraint of a common fit parameter $\sigmaAstat$
in
\be
\Sigma^\text{stat}_{\text{A,}i}(u,a/L)=
\sigmaAstat(u)+\rho'_i(u)\times(a/L)^2 \;.
\label{extrap_cst}
\ee
Fit results from two examples of fits of the first kind are deferred 
to \tabs{tab:sigastatsepfit1_res} and \ref{tab:sigastatsepfit2_res} 
in \App{App_results}.
These fits employ the ansatz (\ref{extrap_sep}), including the data at all 
available lattice resolutions in one case, whereas in the other case the 
coarsest data, $L/a=6$, are discarded and the fit ansatz is just a constant 
($\rho\equiv 0$).
The nice consistency of the two sets of results provides already clear
evidence for the very weak overall lattice spacing dependence of the step
scaling functions $\Sigma^\text{stat}_{\text{A,}i}(u,a/L)$, particularly 
beyond $L/a=6$, which can also be inferred from \tabs{tab:zastat1_res} 
and \ref{tab:zastat2_res} by direct inspection of the raw data. 

Our final extrapolation to the continuum is based on our experiences gained 
from the quantitative investigations of the running of the QCD coupling and 
quark masses in the SF scheme \cite{alpha:Nf2_2,msbar:Nf2}, where the cutoff 
effects of the corresponding step scaling functions were found to be very
small as well.
The continuum limits were thus obtained according to \eq{extrap_cst} by 
fitting the two values of $\Sigma^\text{stat}_{\text{A,}i}$ on the finer 
lattices ($L/a=8,12$) simultaneously for 
$i\in\{\text{s},\text{HYP1},\text{HYP2}\}$ to a common constant 
$\sigmaAstat$ ($\rho'\equiv 0$), separately for each coupling $u$.
We then added linearly the difference between the fit and the $L/a=8$ 
result as a systematic error.
These constrained constant fits are displayed in \fig{fig:Sigma_zastat}, 
and the resulting pairs of $u$ and continuum values $\sigmaAstat(u)$ are 
summarized in \tab{tab:sigastat_res}.
By explicitly trying various different extrapolations along the ans\"atze 
(\ref{extrap_sep}) and (\ref{extrap_cst}) (while not only quadratic but
also linear in $a$) we verified with confidence that, within the achievable 
statistical accuracy, our data do not show any significant dependence on 
the lattice spacing and that the continuum limit is well under control.
\subsection{Non-perturbative scale evolution}
%
\FIGURE[t]{
\epsfig{file=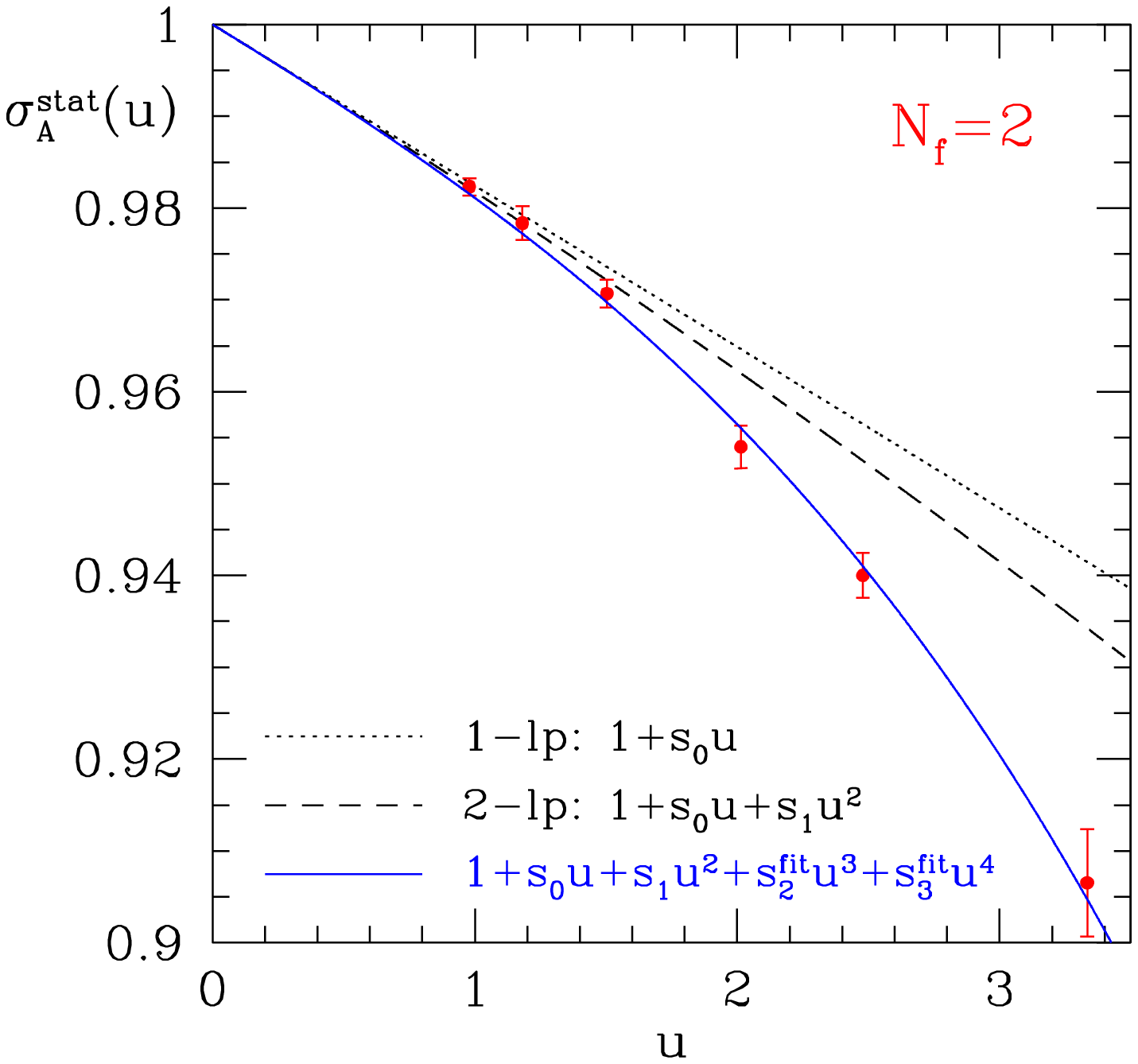,width=10.0cm}
\caption{
Continuum step scaling function $\sigmaAstat(u)$ and its polynomial fit.
}\label{fig:sigma_zastat}
}
%
For the next steps of the analysis it is most convenient to represent the
continuum step scaling function $\sigmaAstat$ as a smooth function of $u$.
To this end, we start from \eq{me_PhiRGI} to write down the expression that
relates $\sigmaAstat(u)$ to the anomalous dimension $\gamma$ and the 
$\beta$--function, namely
\be
\sigmaAstat(u)=
\exp\left\{-\int_{\sqrt{u}}^{\sqrt{\sigma(u)}} \rmd g\,
\frac{\gamma(g)}{\beta(g)}\right\} \;,
\label{SSF_cont_pert}
\ee 
where $\sigma(u)$ is the step scaling function of the coupling, determined 
by \cite{alpha:Nf2_2}
\be
-2\ln(2)=
\int_u^{\sigma(u)}\rmd x\,\frac{1}{\sqrt{x}\,\beta(\sqrt{x})} \;. 
\label{SSF_coupling}
\ee
The first non-universal coefficients in the perturbative expansions of 
$\gamma$ and $\beta$ are given for the SF scheme by the two-loop result
$\gamma_1$ as already quoted in (\ref{gam_1}) and the three-loop expression
(cf.~\cite{alpha:Nf2_2})
\be
b_2=
\frac{1}{(4\pi)^3}\,\Big(\,0.483-0.275\Nf+0.0361\Nf^2-0.00175\Nf^3\,\Big)\;.
\ee
These formulae imply that $\sigmaAstat$ has a perturbative expansion of the 
form
\be
\sigmaAstat(u)=
1+s_0u+s_1u^2+s_2u^3+\ldots
\label{SSF_cont_fit}
\ee
with, for instance, the two leading coefficients found to be
\bea
s_0 
& = &
\ln(2)\gamma_0 \;, \\
s_1 
& = &
\half s_0^2+s_0\ln(2)b_0+\ln(2)\gamma_1 \;.
\eea

Guided by \eq{SSF_cont_fit}, we therefore represent the non-perturbative 
results for $\sigmaAstat(u)$ in \tab{tab:sigastat_res} (with added 
statistical and systematic errors) by an interpolating fit to an ansatz 
polynomial in $u$, where $s_0,s_1$ are restricted to their perturbative 
predictions and up to three additional free fit parameters are allowed for.
All of these fits describe the data well, and we quote the two-parameter 
fit (the curve of which is shown in \fig{fig:sigma_zastat}) as the final 
representation for the functional form of $\sigmaAstat(u)$.
The stability of the polynomial fits was further checked by fits with only 
$s_0$ (or even no coefficient at all) constrained to its perturbative 
value, which led to compatible results, including a fit to
$(\sigmaAstat(u)-1)/u=s_0+s_1u+\ldots$ to reproduce the perturbative 
prediction for $s_0$.

Moreover, \fig{fig:sigma_zastat} demonstrates --- in contrast to previous 
calculations of the non-perturbative scale evolution within the SF scheme 
of other renormalized observables in quenched and two-flavor 
QCD \cite{zastat:pap3,msbar:pap1,alpha:Nf2_2,msbar:Nf2} --- that the 
present case of the $\Nf=2$ static axial current provides another\footnote{
Comparable deviations between perturbative and non-perturbative running
have so far only been observed for some of the SF schemes studied in the
renormalization of four-fermion operators \cite{bk:pap1}.
} 
example for a significant deviation of the non--perturbative data from the 
perturbative behaviour, which even sets in already at moderate couplings 
$u$.

Now we use $\sigmaAstat(u)$ given by the fit function in order to solve 
the following joint recursion to evolve the coupling and the renormalized 
matrix element $\Phi$ from a low-energy scale $1/\Lmax$ implicitly defined 
by
\be
u_0=\gbar^2(\lmax)=4.61
\ee
to the higher energy scales $1/L_k$, $k=0,1,\ldots,8$ 
(with $L_0\equiv\lmax$):
\bea
u_0=4.61 \;,\quad \sigma(u_{k+1})=u_k
& \quad\Rightarrow\quad &
u_k=\gbar^2(L_k) \;,\quad L_k=2^{-k}\lmax \;, 
\label{recursion_sig} \\
v_0=1 \;,\quad v_k=\left[\,\prod_{i=1}^k\sigmaAstat(u_i)\,\right]^{-1}
& \quad\Rightarrow\quad &
v_k=\frac{\Phi(1/\lmax)}{\Phi(1/L_k)} \;.
\label{recursion_siga}
\eea
For this purpose, also the scale evolution of the coupling is
parameterized by an interpolating polynomial of the form
$\sigma(u)=u+\sigma_0u^2+\sigma_1u^3+\ldots\,$, for which the exact results 
of the corresponding fit (and its covariance matrix) were available
from~\cite{msbar:Nf2,msbar:Nf2_ssffit}.
Since the errors on the step scaling functions stem from different 
simulation runs and are hence uncorrelated, the errors on the fit 
parameters in the polynomials for $\sigma(u),\sigmaAstat(u)$ and those on 
the recursion coefficients $u_k,v_k$ calculated from them can be estimated 
and passed through the recursion straightforwardly by the standard error 
propagation rules.
\clearpage
\subsection{The universal renormalization factor}
%
\TABLE{
\centering
\renewcommand{\arraystretch}{1.25}
\begin{tabular}{ccclcc} 
\toprule
  & & & \multicolumn{3}{c}{$\PhiRGI/\Phi(\lmax^{-1})$} \\ \cmidrule(lr{.75em}){4-6}
$k$ & $u$ & & \clmc{2/3-loop} & & \clmc{1/2-loop} \\ 
\midrule
  0 & 4.610 & & 0.853    & & 0.851 \\
  1 & 3.032 & & 0.863(3) & & 0.862 \\
  2 & 2.341 & & 0.871(5) & & 0.870 \\
  3 & 1.918 & & 0.875(6) & & 0.874 \\
  4 & 1.628 & & 0.878(6) & & 0.877 \\
  5 & 1.414 & & 0.879(7) & & 0.878 \\
  6 & 1.251 & & 0.880(7) & & 0.879 \\
  7 & 1.121 & & 0.880(7) & & 0.879 \\
  8 & 1.017 & & 0.880(7) & & 0.880 \\ \bottomrule
\end{tabular}
\caption{
Evaluation of \eq{Phiratio}, exploiting the perturbative knowledge of the
$\gamma$-- and $\beta$--functions in the SF scheme.
}\label{tab:phiratio_res}
}
%
\FIGURE[t]{
\epsfig{file=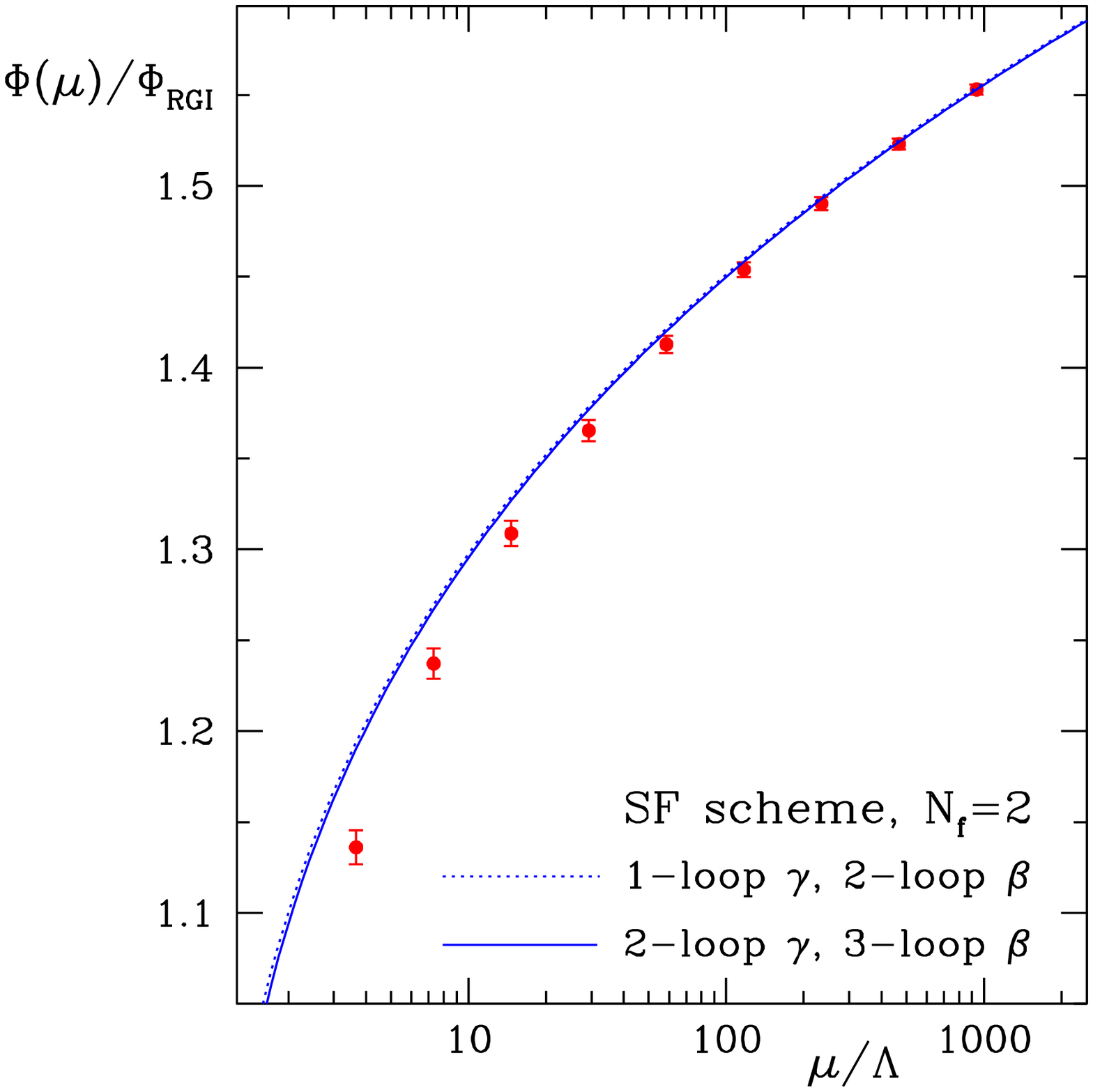,width=10.0cm}
\caption{
Numerically computed values of the running matrix element of the static 
axial current in the SF scheme compared to perturbation theory.
The dotted and solid lines are obtained from \eq{me_PhiRGI} using the
1/2-- and 2/3--loop expressions for the $\gamma$-- and $\beta$--functions,
respectively, as well as $\ln(\Lambda\Lmax)=-1.298(58)$ from 
\protect\Ref{alpha:Nf2_2}.
}\label{fig:PhiSF_stat}
}
%
Finally, by virtue of \eqs{me_PhiRGI} and (\ref{recursion_siga}), 
we proceed to relate the renormalized matrix element
$\Phi(\mu)=\zastat(L)\Phibare(g_0)$, $\mu=1/L$, at $L=\lmax$ to the RGI 
one as 
\bea
\frac{\PhiRGI}{\Phi(1/\lmax)}
& = &
v_k^{-1}\,\frac{\PhiRGI}{\Phi(1/L_k)}
\label{Phiratio} \\
& = &
{\zastat(2^{-k}\lmax)\over\zastat(\lmax)}\,
\left[\,2b_0\gbar^2(\mu)\,\right]^{-\gamma_0/(2b_0)}
\exp\left\{-\int_0^{\gbar(\mu)} \rmd g 
\left[\,{\gamma(g)\over\beta(g)}-{\gamma_0 \over b_0 g}\,\right]
\right\} \nonumber
\eea
with $\mu=2^k/\lmax$ and the ratios of $Z$--factors, $v_k$, to be taken 
from the non-perturbative solution of the recursion (\ref{recursion_siga})
discussed in the foregoing subsection.
After numerical integration of the second factor, $\PhiRGI/\Phi(1/L_k)$,
with $\gbar^2=u_k$ by employing the perturbative expressions for the 
$\gamma$-- and $\beta$--functions at two- and three-loop accuracy, 
respectively, we arrive at the series of numbers listed 
in~\tab{tab:phiratio_res}.
For $k\ge 3$ they show a remarkable stability in the coupling\footnote{
The deviation in the case $k=0$ is due to the difference between the
perturbative and the non-perturbative values of $\sigma(u)$ at large $u$
(see \cite{alpha:Nf2_2}).
}
$u_k$ such that we select $k=6$ --- yielding
\be
\frac{\zastat(\lmax)}{\zastat(L)}=0.762(5)
\quad \mbox{at} \quad L=2^{-6}\lmax
\label{Zratio_res}
\ee
and the coupling value ($u_6$) entering the second factor 
in \eq{Phiratio} to lie safely in the perturbative regime --- in order to 
obtain our central result
\be
\frac{\PhiRGI}{\Phi(\mu)} = 0.880(7)
\quad \mbox{or} \quad
\frac{\Phi(\mu)}{\PhiRGI} = 1.137(9)
\quad \mbox{at} \quad \mu=1/\lmax
\label{Phiratio_res}
\ee
in the SF scheme\footnote{
Still, for the perturbative error in this step to be negligible, it is 
crucial that $\gamma$ is known to two-loop precision and $\beta$ to 
three-loop. 
}.
Hence, through this analysis we have succeeded in connecting the low- and 
high-energy scales $L=\lmax$ and $L=2^{-6}\lmax$ non-perturbatively as
intended.

Note that, since the continuum limit has been taken, any regularization 
dependence has been removed from the result (\ref{Phiratio_res}) so that 
ideally the error on $\Phi(\mu)/\PhiRGI$ of about 0.8\% should only be
included in estimates on matrix elements in the continuum limit.

In \fig{fig:PhiSF_stat} we compare the numerically computed running 
with the corresponding cur\-ves in perturbation theory.
For the argument $\mu/\Lambda=1/(L_k\Lambda)$, $k=0,1,\ldots,8$, we plot 
the points $\Phi(1/L_k)/\PhiRGI$ calculated from (\ref{Phiratio}) using the 
universal result (\ref{Phiratio_res}).
Here, the physical scale $\Lambda$ is implicitly determined by
$\ln(\Lambda\Lmax)=-1.298(58)$ resulting from the 
recursion (\ref{recursion_sig}) \cite{alpha:Nf2_2,msbar:Nf2}, and the errors 
of the points in \fig{fig:PhiSF_stat} come from the coefficients $v_k$.
As expected from the behaviour of $\sigmaAstat(u)$ 
in~\fig{fig:sigma_zastat}, this comparison between the non-perturbative and
perturbative running again reveals that good agreement with the 
perturbative approximation is only observed at high scales, while the 
difference grows up to 5\% towards smaller energies of the order
$\mu\approx4\Lambda$.

\section{$\zastat$ at the low-energy matching scale}
\label{Sec_match}
%
\TABLE{
\centering
\renewcommand{\arraystretch}{1}
\def\onelp{\small{1-lp}}
\begin{tabular}{ccclccccl} \toprule
 $\beta$  & $\kappa$  & $L/a$ & $ \gbar^2(L)$ & &  $Z^\text{stat}_\text{A,s}$ &
     $Z^\text{stat}_\text{A,HYP1}$ & $Z^\text{stat}_\text{A,HYP2} $ & $\castat$ \\ \midrule
 5.20 & 0.13600 & 4 & 3.65(3)   & & 0.84621(63) & 0.84550(53) & 0.88820(58) & \onelp \\
      &         &   &           & & 0.84846(63) & 0.85760(54) & 0.95061(65) & 0 \\[0.5ex]
      &         & 6 & 4.61(4)   & & 0.79204(58) & 0.79631(52) & 0.84456(52) & \onelp  \\
      &         &   &           & & 0.79409(58) & 0.80730(52) & 0.90004(55) & 0 \\\cmidrule(lr){1-9}  
 5.29 & 0.13641 & 4 & 3.394(17) & & 0.85323(60) & 0.85218(50) & 0.89202(55) & \onelp  \\
      &         &   &           & & 0.85541(60) & 0.86385(51) & 0.95227(61) & 0 \\[0.5ex]
      &         & 6 & 4.297(37) & & 0.79954(69) & 0.80384(60) & 0.85006(59) & \onelp  \\
      &         &   &           & & 0.80152(69) & 0.81444(61) & 0.90358(62) & 0 \\[0.5ex]
      &         & 8 & 5.65(9)   & & 0.75464(71) & 0.76110(65) & 0.80934(65) & \onelp  \\
      &         &   &           & & 0.75653(71) & 0.77124(65) & 0.86058(67) & 0 \\\cmidrule(lr){1-9} 
 5.40 & 0.13669 & 4 & 3.188(24) & & 0.86090(63) & 0.85965(53) & 0.90020(56) & \onelp  \\
      &         &   &           & & 0.86302(63) & 0.87103(53) & 0.95911(62) & 0 \\[0.5ex]
      &         & 6 & 3.864(34) & & 0.81111(68) & 0.81448(60) & 0.85700(59) & \onelp  \\
      &         &   &           & & 0.81300(68) & 0.82458(60) & 0.90774(63) & 0 \\[0.5ex]
      &         & 8 & 4.747(63) & & 0.77310(64) & 0.77816(59) & 0.82316(57) & \onelp  \\
      &         &   &           & & 0.77491(65) & 0.78785(59) & 0.87185(60) & 0 \\ \bottomrule
\end{tabular}   
\caption{
Results for $\zastat$ with $\ct$ set to its 2--loop value, both for 
$\castat={\rm 1-loop}$ and $\castat=0$. 
The values of $\gbar^2$ are from \cite{alpha:Nf2_2}.
The hopping parameters $\kappa$ used in the simulations are taken to be the 
critical ones ($\kappa_c$) of \cite{Gockeler:2004rp}.
}\label{tab:zastatmatch_res}
}
%
\TABLE{
\centering
\renewcommand{\arraystretch}{1}
\def\onelp{\small{1-lp}}
\begin{tabular}{cccclclcl} \toprule
 $\Sstat$ & & $\beta$ & & \clmc{$\zastat$} & & \clmc{$\ZPhi$} 
& & $\castat$ \\ \midrule
 s    & & 5.20 & &  0.7920(6)   & &  0.6970(6)(56)   & &  \onelp  \\
      & &      & &  0.7941(6)   & &  0.6988(6)(56)   & &  0       \\[0.75ex]
      & & 5.29 & &  0.7873(28)  & &  0.6928(28)(55)  & &  \onelp  \\
      & &      & &  0.7892(28)  & &  0.6945(28)(56)  & &  0       \\[0.75ex]
      & & 5.40 & &  0.7784(28)  & &  0.6850(28)(55)  & &  \onelp  \\
      & &      & &  0.7802(28)  & &  0.6866(28)(55)  & &  0       \\\midrule
 HYP1 & & 5.20 & &  0.7962(5)   & &  0.7007(5)(56)   & &  \onelp  \\
      & &      & &  0.8073(5)   & &  0.7104(5)(57)   & &  0       \\[0.75ex]
      & & 5.29 & &  0.7922(27)  & &  0.6971(27)(56)  & &  \onelp  \\
      & &      & &  0.8026(27)  & &  0.7063(27)(57)  & &  0       \\[0.75ex]
      & & 5.40 & &  0.7832(26)  & &  0.6892(26)(55)  & &  \onelp  \\
      & &      & &  0.7930(27)  & &  0.6978(27)(56)  & &  0       \\\midrule
 HYP2 & & 5.20 & &  0.8446(5)   & &  0.7432(5)(59)   & &  \onelp  \\
      & &      & &  0.9000(5)   & &  0.7920(5)(63)   & &  0       \\[0.75ex]
      & & 5.29 & &  0.8390(25)  & &  0.7383(25)(59)  & &  \onelp  \\
      & &      & &  0.8919(26)  & &  0.7849(26)(63)  & &  0       \\[0.75ex]
      & & 5.40 & &  0.8279(24)  & &  0.7286(24)(58)  & &  \onelp  \\
      & &      & &  0.8769(26)  & &  0.7717(26)(62)  & &  0       \\ \bottomrule
\end{tabular}
\caption{
Results for $\zastat$ and $\ZPhi$ for three bare gauge coupling values
corresponding to our low-energy matching point at $\gbsq=4.61$ in the SF
scheme, distinguishing the three static discretizations used.
}\label{tab:zastattotal_res}
}
%
Connecting a bare matrix element of the static-light axial current to the 
RGI one according to \eq{PhiRGI_ZRGI} amounts to multiply the bare lattice 
operator with the total renormalization factor
\be
\ZPhi(g_0)\equiv
\left.\frac{\PhiRGI}{\Phi(\mu)}\,\right|_{\,\mu=1/\lmax}\,\times\,
\left.\zastat(g_0,L/a)\,\right|_{\,L=\lmax} \;,
\label{ZRGI}
\ee
which involves --- in addition to the universal ratio $\PhiRGI/\Phi(\mu)$, 
\eq{Phiratio_res} --- the value of the renormalization factor 
$\zastat(g_0,L/a)$ at the finite, low-energy renormalization scale $\lmax$ 
implicitly fixed by the condition $\gbsq(\lmax)=4.61$ in the intermediate SF 
scheme.

Following the steps of the analogous computation in the case of the running
quark mass in QCD \cite{msbar:Nf2}, we now derive the second factor 
in \eq{ZRGI} for a few values of the lattice spacing or the bare coupling, 
respectively.
As pointed out before, this contribution is non-universal, and in the form 
given it will be valid only for our static-light actions, consisting of 
non-perturbatively improved Wilson fermions with plaquette gauge action and 
$\csw$ as specified in \cite{impr:csw_nf2} characterizing the light quark 
sector and the three discretizations s, HYP1 and HYP2 employed for the 
static quark flavour.
For $\castat$ we insert the one-loop values recently determined for these
static actions in \cite{HQET:statprec} and reproduced in~\App{App_lattice}. 
It thus remains to compute $\zastat(g_0,\lmax/a)$ for the values 
$\beta=5.2,5.29,5.4$, which lie well within the range of bare couplings 
commonly used in simulations of two-flavour QCD in physically large volumes.
The associated simulation parameters and results are summarized 
in \tab{tab:zastatmatch_res}.
In order to allow for studying the influence of $\castat$ on future
continuum extrapolations of renormalized matrix elements, we also provide
estimates for $\zastat(g_0,L/a)$ with $\castat$ being set to zero.

While one of the simulations at the largest bare coupling is exactly at the 
target value for $\gbar^2$, the two other series of simulations require a 
slight interpolation.
This has been done adopting a fit ansatz motivated by \eq{me_PhiRGI},
\be
\ln\left(\zastat\right)=c_1+c_2\ln(\gbar^2) \;,
\label{logfit_zastat}
\ee
to interpolate $\zastat$ between those two values of $\gbar^2$ straddling 
the target value $4.61$, whereby the fit takes into account the 
(independent) errors of both $\zastat$ and $\gbar^2$.
We then augmented the fit error by the difference between the fit result 
via \eq{logfit_zastat} and the result from a simple two-point linear 
interpolation in $\gbar^2$.
The values of the coefficient $c_2$ in the fit (\ref{logfit_zastat}) are 
found to deviate not more than by about $0.03$ (with errors on the 7\% 
level) from $\gamma_0/(2b_0)=-6/29\approx -0.2069$.

The resulting numbers for $\zastat(g_0,\lmax/a)$ and finally for the total
renormalization factor $\ZPhi(g_0)$, cf.~\eq{ZRGI}, are collected 
in \tab{tab:zastattotal_res}.
The first error on $\ZPhi$ stems from the error of $\zastat$, whereas the
second one embodies the 0.8\% uncertainty in the universal factor 
$\PhiRGI/\Phi$ and, provided that the renormalized matrix element of the 
static current is available at several lattice spacings, should not be 
added in quadrature to the error on the latter before the continuum limit 
has eventually been taken.
For later use, a representation of the numerical results 
of~\tab{tab:zastattotal_res} by interpolating polynomials can be found 
in~\tab{tab:zastatpoly_res} in~\App{App_results}.
Comparing the cases $\castat={\rm 1-loop}$ and $\castat=0$, we observe
that only for the static action HYP2 the change in the renormalization
factors is about 6\% and thereby non-negligible, which however can be 
attributed to the fact that for this action the one-loop coefficient of 
$\castat$ is by one order of magnitude larger than for the static 
discretization HYP1 (cf.~\eq{ca_stat_new}).

\section{Matrix elements at finite values of the quark mass} 
\label{Sec_MEfinite}
It was already outlined in the introduction that in order to employ results
from the static effective theory, one has to translate its RGI matrix 
elements to those in QCD at finite values of the heavy quark mass.
For the special case of the matrix element of the axial current between the 
vacuum and the heavy-light pseudoscalar, this conversion to the so-called 
`\emph{matching scheme}' amounts to a multiplication with a function 
$\Cps(M/\lMSbar)$, viz.
\be
\Fps\sqrt{\mps}=
\Cps\left(M/\lMSbar\right)\PhiRGI=
\Cps\left(M/\lMSbar\right)\ZPhi
\ketbra{\,{\rm PS}\,}{\,\Astat\,}{\,0\,}
\label{me_QCD_RGI}
\ee
up to $\Or(1/M)$ corrections, where it is theoretically as well as 
practically advantageous to express $\Cps$ in terms of a ratio of RGIs 
as \cite{zastat:pap3,HQET:pap3}
\be
\Cps\left(M/\lMSbar\right)=
\left[\,2b_0\gbarMSbar^2\,\right]^{\gamma_0/(2b_0)}
\exp\left\{\int_0^{\gbarMSbar} \rmd g 
\left[\,{\gammatch(g)\over\beta(g)}-{\gamma_0 \over b_0 g}\,\right]
\right\} \;;
\ee
$\gammatch$ denotes the anomalous dimension of the current in the matching 
scheme. 
As we will see in the next subsection, $\Cps$ is very well under control in
perturbation theory.

To exploit~\eq{me_QCD_RGI} in order to determine the decay constant $\Fps$, 
after having non-perturbatively solved the renormalization problem of the 
static axial current, is the main purpose of this section.
\subsection{Perturbative evaluation of the conversion function}
\label{Sec_MEfinite_Cps}
%
\FIGURE[t]{
\epsfig{file=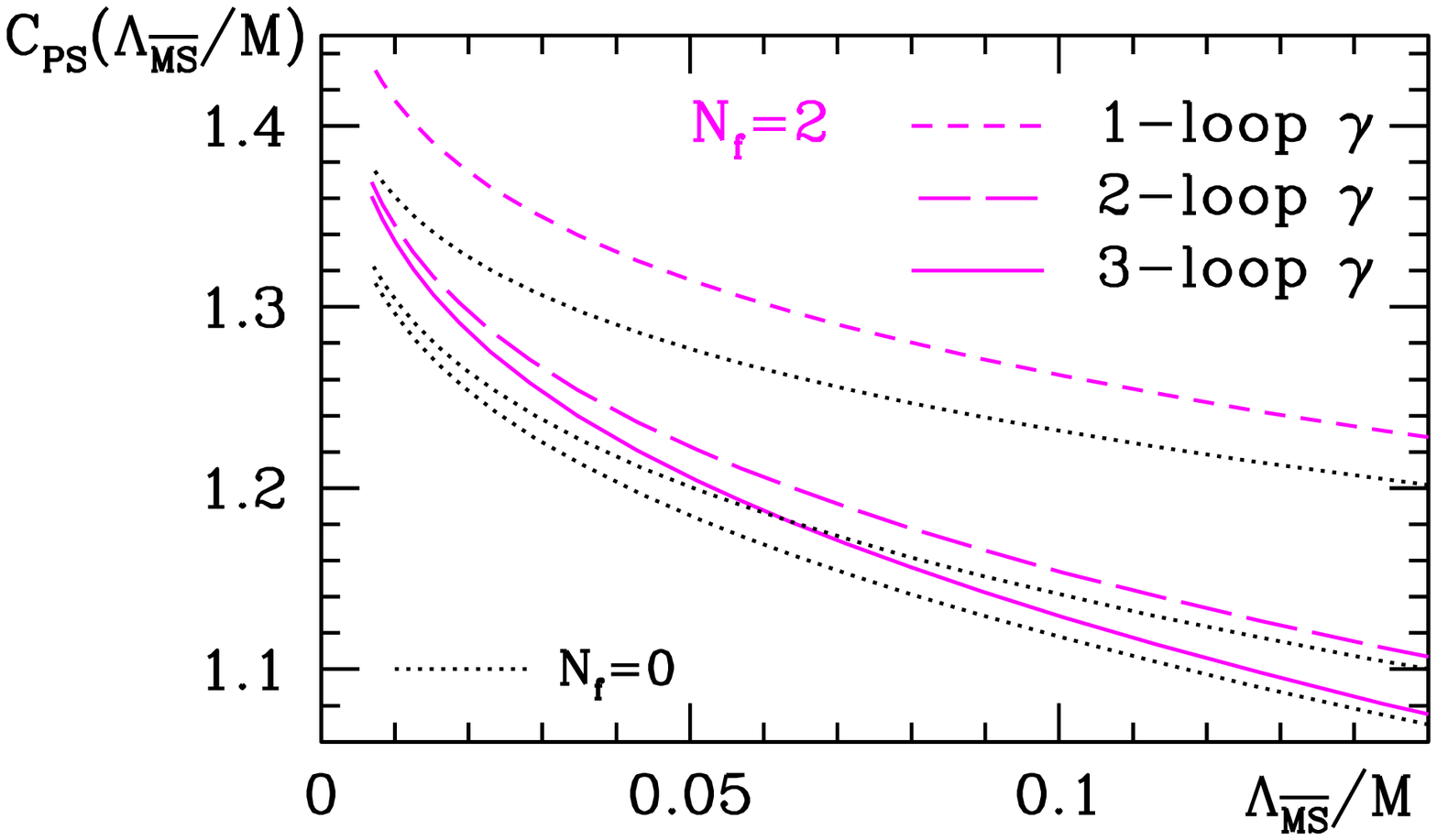,width=11.0cm}
\caption{
$\Nf=2$ conversion factor to the matching scheme, which translates the RGI
matrix element of $\Astat$ to the one at finite mass.
A continuous parameterization using the anomalous dimension $\gamma$ to
two- and three-loop accuracy, respectively, is given in \eq{cpsfit_nf2}.
The dotted curves show the corresponding conversion function for
$\nf=0$ \cite{zastat:pap3,HQET:pap3} for comparison.
}\label{fig:CpsMatch}
}
%
The numerical evaluation of the perturbative approximation of the
conversion function $\Cps(M/\Lambda_{\rm \overline{MS}})$ has been explained 
in detail in appendix~B of \Ref{HQET:pap3}.
This analysis is straightforwardly extended to the present case of
two-flavour QCD, $\Nf=2$:
Using the anomalous dimension in the matching scheme that involves
(i) the anomalous dimension of the corresponding effective theory operator
in the $\MS$ scheme up to three-loop
order \cite{ChetGrozin,Ji:1991pr,BroadhGrozin1,Gimenez:1992bf} and
(ii) matching coefficients between the effective theory and physical QCD
operator up to two
loops \cite{stat:eichhill1,stat:eichhill_za,BroadhGrozin2} --- together
with the four-loop $\beta$--function \cite{vanRitbergen:1997va}
and quark mass anomalous
dimension \cite{Chetyrkin:1997dh,Vermaseren:1997fq} in the $\MS$ scheme ---,
the numerically evaluated conversion function
$\Cps(M/\Lambda_{\rm \overline{MS}})$ is shown in \fig{fig:CpsMatch}.

As in the quenched case \cite{zastat:pap3,HQET:pap3} it is actually more
convenient to represent $\Cps$ as a continuous function in terms of the
variable
\be
x\equiv
\frac{1}{\ln\left(\M/\Lambda_{\msbar}\right)}
\label{cpsarg}
\ee
in a form that is motivated by \eq{me_PhiRGI}.
This results in
\bea
\Nf=2:\quad
\Cps(x)=
\left\{
\begin{array}{ll}
x^{\gamma_0/(2b_0)}\left\{\,1-0.107\,x+0.093\,x^2\,\right\}
& \;\; \mbox{2--loop $\gamma$}\\
\\
x^{\gamma_0/(2b_0)}
\left\{\,1-0.118\,x-0.010\,x^2+0.043\,x^3\,\right\}
& \;\; \mbox{3--loop $\gamma$}
\end{array}
\right. \,,
\label{cpsfit_nf2}
\eea
with $b_0=(11-\frac{2}{3}\Nf)/(4\pi)^2$ and $\gamma_0=-1/(4\pi^2)$,
whereby the prefactor encodes the leading asymptotics as $x\rightarrow0$.
For 3--loop $\gamma$, the precision of the
parameterization (\ref{cpsfit_nf2}) is at least 0.2\% for $x\leq 0.6$, and
one may attribute an error of at most 2\% owing to the perturbative
approximation underlying this determination of the conversion function.

For future purposes, we also quote the corresponding parameterization for
the three-flavour theory:
\bea
\Nf=3:\quad
\Cps(x)=
\left\{
\begin{array}{ll}
x^{\gamma_0/(2b_0)}\left\{\,1-0.144\,x+0.130\,x^2\,\right\}
& \;\; \mbox{2--loop $\gamma$}\\
\\
x^{\gamma_0/(2b_0)}
\left\{\,1-0.161\,x+0.062\,x^2+0.007\,x^3\,\right\}
& \;\; \mbox{3--loop $\gamma$}
\end{array}
\right. \,.
\label{cpsfit_nf3}
\eea
\subsection{Application: non-perturbative renormalization of $\Fbsstat$}
\label{Sec_MEfinite_appl}
For an immediate use of the results obtained, we present a first 
non-perturbative computation of the decay constant $F_{\rm B_{\rm s}}$ in 
the static limit, based on data for the bare matrix element of $\Astat$ in
large volume and for the static action $S^\text{HYP2}\lh$.
To this end we have used one of the sets of unquenched (two degenerate 
flavours) configurations produced by the ALPHA Collaboration for setting the 
scale for the $\Nf=2$ $\Lambda$--parameter and RGI quark 
mass \cite{alpha:Nf2_2,msbar:Nf2} by simulations in physically large 
volumes~\cite{fbstat:Nf2,Nf2SF:autocorr}.
The bare parameters are given by $\beta=5.3$, $\kappa=0.1355$, 
$V=24^3\times 32$, and the RGI light quark mass \cite{msbar:Nf2} turns out 
to be $171(4)\,\MeV$. 
For the conversion to MeV we have used $a=0.078(1)\,\Fm$ at $\beta=5.3$ as 
given in \Ref{cern:Nf2light_1}.
The value of the quark mass is at the upper end of the result for $M_{\rm s}$ 
in~\cite{msbar:Nf2} (i.e.~$M_{\rm s}=138(31)\,\MeV$). 
Anyway, the dependence of $F_{\rm B_{\rm s}}$ on the exact value of $M_{\rm s}$ 
is expected to be quite mild, as the JLQCD Collaboration, for instance, has 
reported ${F_{\rm B_{\rm s}}}/{F_{\rm B_{\rm u}}}=1.13(3)(^{+13}_{-2})$ 
in~\Ref{Aoki:2003xb}.\footnote{
A consistent, albeit less precise number was also found 
in~\Ref{fbstat:ukqcd} at a finite lattic spacing corresponding 
to $\beta=5.2$.
}
According to that we would have to correct our result on $F_{\rm B_{\rm s}}$ 
by decreasing it by about $3\%$. 
Such a correction is however below the $5\%$ statistical error we are able 
to achieve at this lattice resolution.

The computational details closely follow what has been done in the quenched 
case discussed in~\Refs{fbstat:pap1,fbstat:pap2}.
To suppress excited B-meson state contributions to the correlation 
functions, we introduce wave functions $\omega({\bf x})$ at the boundaries 
of the SF such that the correlators $f_{\rm A}^{\rm stat}$ and $f_1^{\rm stat}$ 
(cf.~eqs.~(\ref{fastat})~--~(\ref{f1hl}) in \sect{Sec_rscheme}) take the 
form
\be
f_{\rm A}^{\rm stat}(x_0,\omega_i)=
-{{1}\over{2}}\left\langle(A_{\rm I}^{\rm stat})_0(x)\, 
{\cal{O}}(\omega_i)\right\rangle \;, 
\quad
f_1^{\rm stat}(\omega_i,\omega_j)=
-{{1}\over{2}}\left\langle{\cal{O}}'(\omega_i)\,
{\cal{O}}(\omega_j)\right\rangle\;,
\ee
with\footnote{
In the spirit of \Ref{BMP}, in \eq{O'O} we replace one of the spatial sums 
by a sum over eight separated points, which in practice is realized in line
with the inversion of the Dirac operator by shifting the source at the
origin into all octants of the spatial volume $L^3$.
}
\be
{\cal O}(\omega)=
{{a^6}\over{L^3}}\sum_{\bf y,z}\overline{\zeta}_{\rm h}({\bf y})\gamma_5
\,\omega({\bf y}-{\bf z})\,\zeta_{\rm l}({\bf z}) \;,
\quad
{\cal O}'(\omega)=
{{a^6}\over{L^3}}\sum_{\bf y,z}\overline{\zeta}_{\rm l}'({\bf y})\gamma_5
\,\omega({\bf y}-{\bf z})\,\zeta_{\rm h}'({\bf z}) \;,
\label{O'O}
\ee
and the improved version $(A_{\rm I}^{\rm stat})_0$ of the static-light axial 
current defined as
\be
(A_{\rm I}^{\rm stat})_0(x)= 
A^{\rm stat}_0(x)+a\,c_{\rm A}^{\rm stat}\delta A_0^{\rm stat}(x) \;.
\ee
We restrict ourselves to a choice of four spatially periodic wave functions
\bea
\omega_i({\bf x})
& = &
\frac{1}{N_i}\sum_{\bf n\,\in\,\mathbb{Z}^3}
\overline{\omega}_i\left(|{\bf x-n}L|\right) \;, 
\quad i=1,\dots,4 \;, 
\nonumber\\[1ex]
\overline{\omega}_1(r)
& = &
a^{\,-3/2}\,\Exp^{\,-r/a} \;,\qquad 
\overline{\omega}_2(r)=a^{\,-3/2}\,\Exp^{\,-r/2a} \;,
\nonumber\\[1ex]
\overline{\omega}_3(r)
& = &
a^{\,-5/2}\,r\,\Exp^{\,-r/2a} \;,\quad 
\overline{\omega}_4(r)=a^{\,-3/2}\,\Exp^{\,-r/4a} \;,
\eea
with the coefficients $N_i$ normalizing them such that
$a^3\sum_{\bf x}\omega_i^2({\bf x})=1$ holds.
Numerically, we have approximated the wave functions using the lowest six 
Fourier components in each spatial (positive and negative) direction. 
That reduces the computational cost for the convolutions required to 
calculate $f_1^{\rm stat}$.

The decay constant is then extracted from the expression for the local RGI
matrix element of the static axial current,
\be
\Phi_{\rm RGI}(x_0,\omega_i)=
-Z_{\rm RGI}\left(1+b_{\rm A}^{\rm stat}am_{\rm q}\right)2L^{3/2}\,
{{f_{\rm A}^{\rm stat}(x_0,\omega_i)}
\over{\sqrt{f_1^{\rm stat}(\omega_i,\omega_i)}
}}\,\Exp^{\,(x_0-T/2)E_{\rm eff}(x_0,\omega_i)} \;,
\label{Phi_decay}
\ee
where for $b_{\rm A}^{\rm stat}$ the one-loop formula for the static
discretization $S^\text{HYP2}\lh$ from~\Ref{HQET:statprec} enters.
The effective energy $E_{\rm eff}$ reads 
\be
E_{\rm eff}(x_0,\omega_i)=
\frac{1}{2a}\ln\left[\,
\frac{\fastat(x_0-a,\omega_i)}{\fastat(x_0+a,\omega_i)}\,\right]
\ee
and is shown in \fig{fig:EeffHYP2} for the wave function $\omega_2$. 
This choice is motivated by the fact that $E_{\rm eff}(x_0,\omega_2)$ 
approaches its plateau value earlier than it is the case for the other wave 
functions (which anyway yield consistent values for $x_0\gtrsim 1.4\,\Fm$).
The corresponding bare decay constant $\Phibare$, i.e.~the quantity 
in \eq{Phi_decay} with $Z_{\rm RGI}$ set to one and $b_{\rm A}$ set to zero, 
is displayed in \fig{fig:FeffHYP2}.
Proper linear combinations built from correlators affiliated to the other
wave functions lead to fully compatible graphs.

%
\FIGURE[t]{
\epsfig{file=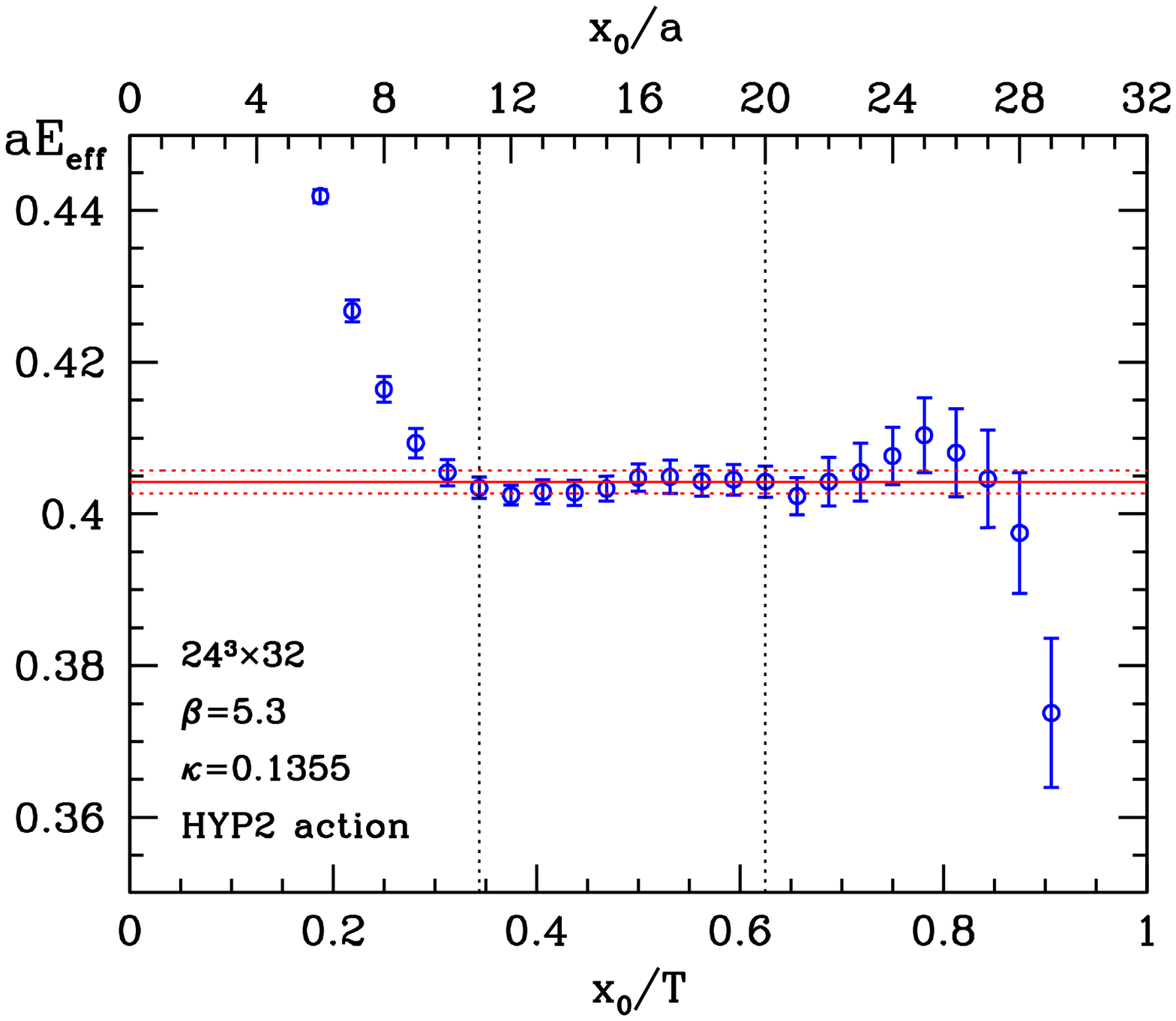,height=10.0cm}
\caption{
Effective energy for the wave function $\omega_2$. 
The plateau region is limited by the dotted vertical lines while the 
horizontal band is the resulting plateau average.
}\label{fig:EeffHYP2}
}
%
\FIGURE[t]{
\hspace*{-0.75cm}
\epsfig{file=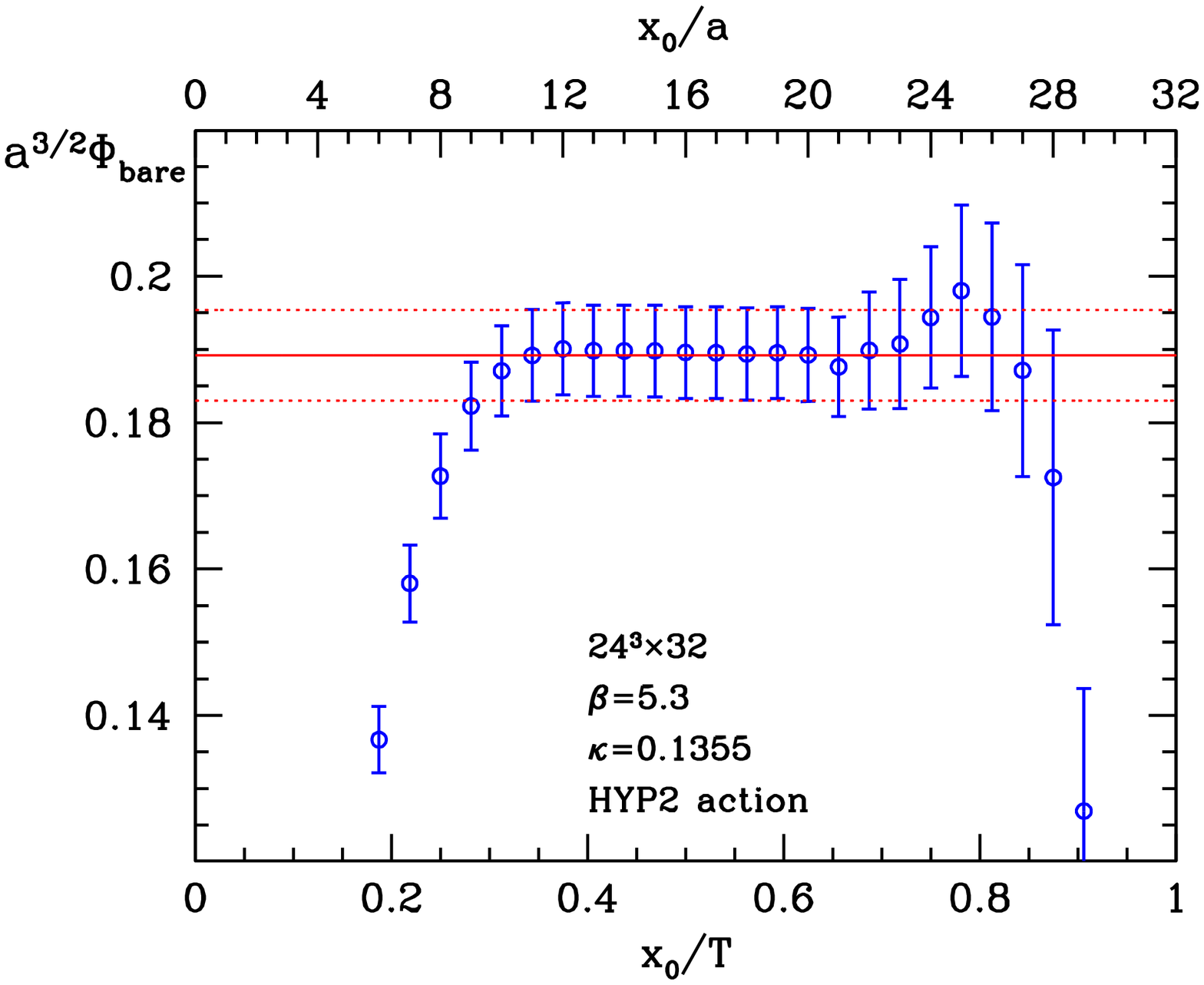,height=10.0cm}
\caption{
Effective \emph{bare} decay constant for the same parameters as 
in~\fig{fig:EeffHYP2}. 
Note that, contrary to the effective energy, in this case the horizontal 
band does not correspond to a plateau average but to the value for 
$\Phibare$ as obtained from the two-state fits of $\fastat(x_0,\omega_2)$ 
(see text).
}\label{fig:FeffHYP2}
}
%
Given the clear plateau in the effective mass, we simply fit it to a 
constant in the region $11\leq x_0/a\leq 20$ and obtain the result
$aE_{\rm stat}=0.4042(15)$.  
Alternatively, we perform a two-state fit of $aE_{\rm eff}(x_0,\omega_2)$ to 
the function $aE_{\rm stat}+b_1\,\Exp^{\,-\Delta^{\rm stat} x_0}$ in the range 
$t_{\rm min}\leq x_0/a\leq 20$, with $t_{\min}\geq 4$. 
The results for $E_{\rm stat}$ from the two fits are in complete agreement.
In addition, in the second case we get numbers for the energy 
gap $a\Delta^{\rm stat}$, which range between $0.5$ and $0.7$ depending on 
the choice for $t_{\rm min}$. 
At the same lattice spacing, the binding energy $aE_{\rm stat}$ from the 
static HYP-actions in the quenched theory turned out to be approximately 
10\% smaller \cite{HQET:mb1m,fbstat:pap2}. 
This is the relative effect one could have guessed by using the one-loop 
expression for the static quark self-energy, considering the shift in 
$\beta$ between the quenched and the $\nf=2$ theory (i.e.~$\beta=6.1$ 
versus $\beta=5.3$, respectively, for $a\approx 0.08\,\Fm$).
Thus, the noise-to-signal ratio of the correlation function, which is 
governed by $aE_{\rm stat}$ and the mass of the lightest 
pseudoscalar \cite{Lepage:1991ui}, is comparable in both cases.

For $\Phibare$, rather than directly fitting it to a constant within the 
plateau region, we first fit the correlation function 
$\fastat(x_0,\omega_2)$ to the two-state ansatz 
$b_2\,\Exp^{\,-E_{\rm stat}x_0}+b_3\,\Exp^{\,-(E_{\rm stat}+\Delta^{\rm stat})x_0}$ 
in order to extract in a second step $\Phibare$ through the ratio of the 
coefficient $b_2$ and the square root of $\fonestat(\omega_2,\omega_2)$, 
as suggested by formula~(\ref{Phi_decay}).
In this way rather precise data could be used in the fit, because the main 
contribution to the statistical error of the effective decay constant comes 
from $\fonestat$.
Moreover, we verified numerically that possible excited state 
contaminations in $\fonestat(\omega_2,\omega_2)$ are strongly suppressed
and hence can be safely neglected.
For varying fit intervals $t_{\rm min}\leq x_0/a\leq 20$ with 
$4\leq t_{\min}\leq 8$, this procedure yields stable numbers that are well
covered by $\Phi_{\rm bare}=0.1892(62)$. 
More details will be provided in a forthcoming 
publication~\cite{fbstat:pap2} extending~\Ref{fbstat:pap1} which, as stated 
above, we follow quite closely here.

Combining the result on $\Phibare$ with~\eq{Phiratio_res}, 
$\zastat(g_0,\lmax/a)|_{\beta=5.3}=0.8382(25)$ (obtained from the 
interpolation quoted at the end of~\App{App_results}) and 
$am_{\rm q}=0.0260(3)$, we get according to~\eq{Phi_decay}:
\be
a^{3/2}\PhiRGI=0.143(5) \;.
\ee
Finally, inserting the proper value of the conversion function
$\Cps(\Mb/\Lambda_{\rm \overline{MS}})=1.24(3)$ (after evaluation of the 
three-loop expression in (\ref{cpsfit_nf2}) from the previous 
subsection\footnote{
In the argument $x$, \eq{cpsarg}, we take $r_0\lMSbar^{(\nf=2)}=0.62(8)$ 
from \cite{alpha:Nf2_2} and the recent $\nf=0$ result for $\Mb$
including $\rmO(1/\mbeauty)$--terms, $r_0\Mb=17.12(22)$ \cite{HQET:mb1m}.
})
into~\eq{me_QCD_RGI}, we arrive at $\Fbsstat=306(14)\,\MeV$, where 
$\mBs=5368(2)\,\MeV$ from~\Ref{PDBook} has been employed.
The quoted error accounts for the statistical uncertainty and the errors of 
the (universal and discretization dependent parts of the) 
renormalization factor as well as for the errors of the lattice spacing and
of~$\Cps$.
It should be recalled, though, that the light quark mass value of the data 
set analyzed and discussed in this section is still slightly above the 
physical strange quark mass.
If the aforementioned decrease by $3\%$ to better meet the strange quark mass
scale is incorporated, our estimate for the $\Bs$-meson decay constant in 
static approximation at a finite lattice spacing of $a\approx 0.08\,\Fm$ 
reads
\be
\Fbsstat=297(14)\,\MeV \;.
\ee
This is broadly compatible with the $\nf=2$ result $\Fbsstat=256(45)\,\MeV$
in the static approximation reported in \Ref{fbstat:ukqcd} by the UKQCD
Collaboration, which however was determined at a coarser lattice spacing
of $a\approx 0.1\,\Fm$ ($\beta=5.2$) and which rests upon the Eichten-Hill 
static quark action and only perturbative renormalization.
 
We can get an idea about discretization effects from the quenched 
($N_{\rm f}=0$) computation~\cite{fbstat:pap2} of the very same quantity,
again using O($a$) improved Wilson fermions and plaquette gauge action but 
the HYP1 action for the static quark. 
There, $\Fbsstat$ \emph{in the continuum} turned out to be $233(18)\,\MeV$,
and in that case the value of $\Fbsstat$ becomes about $10\%$ larger at
$a\approx 0.08\,\Fm$ than the continuum limit \cite{fbstat:pap2}.
Therefore, our result indicates an increase in the two-flavour theory 
either for the value of $\Fbs$, as already observed in~\Ref{lat06:bphys}, 
or for the $\rmO(1/\mbeauty)$--corrections to the decay constant.

\section{Conclusions}
\label{Sec_concl}
We have presented a solution of the scale dependent renormalization problem 
of the static axial current in two-flavour QCD by means of a fully 
non-perturbative computation of the renormalization group running of 
arbitrary renormalized matrix elements $\Phi$ of $\Astat$ in the 
Schr\"odinger functional scheme.\footnote{
As a consequence of the heavy quark spin symmetry that is exact on the 
lattice and owing to the chiral symmetry of the continuum theory,
our computation even yields the scale dependence of all static-light 
bilinears $\lightb\Gamma\heavy$, up to a scale independent, relative
renormalization \cite{zastat:pap3}.
}
In particular, the renormalization factor $\PhiRGI/\Phi(\mu)$ relating the 
matrix element at a specified low-energy scale $\mu$ with the associated RG 
invariant in the continuum limit, \eq{Phiratio_res}, is obtained with a 
good numerical precision, which is comparable with the corresponding study 
in the quenched approximation~\cite{zastat:pap3}.
This is an important prerequisite for a controlled determination of $\Fb$ 
in the static limit of the two-flavour theory.

The use of the alternative discretization schemes 
of~\Refs{fbstat:pap1,HQET:statprec} for the static quark instead of the 
traditional Eichten-Hill action has not only led to a substantial reduction
of the statistical errors of the static-light correlation functions 
involved in our computation, but also entails a convincing universality
test of the continuum limit of the current's lattice step scaling function
(cf.~\fig{fig:Sigma_zastat}).
Similar to the case of the renormalized quark mass 
in $\nf=2$ \cite{msbar:Nf2}, we find this function to be nearly independent 
of the lattice spacing for $a/L<1/6$ and hence conclude that O($a$) 
improvement is very successful also at large values of the SF coupling.
Contrary to the running of the quark mass, however, the scale evolution of
the non-perturbative step scaling function of the renormalized static-light
axial current, \fig{fig:sigma_zastat}, exhibits a pronounced deviation 
from perturbation theory already at moderate couplings.

As a first application, we have combined our non-perturbative 
renormalization factors with a numerical result for the bare matrix element
of $\Astat$ extracted from a large-volume $\nf=2$ simulation, in order to 
estimate the $\Bs$-meson decay constant in the static limit for one value 
of the lattice spacing, $a\approx 0.08\,\Fm$.
In QCD with dynamical quarks, the static approximation may even be the 
starting point of the most viable approach to determine $\Fb$, especially if 
such a computation is supplemented with data on the heavy-light decay 
constant from the charm quark mass region (as already demonstrated in the 
quenched case~\cite{fbstat:pap1,fbstat:pap2}) or with $1/M$--corrections.
For a controlled inclusion of the latter, also the matching between the
effective theory and QCD inherent in the conversion function $\Cps$
(cf.~\eq{me_QCD_RGI}) should then be performed 
\emph{non-perturbatively}~\cite{reviews:NPRrainer_nara}.
A general framework for this is provided by the non-perturbative 
formulation of HQET exposed in~\Ref{HQET:pap1}.
Other approaches, such as the method of heavy quark mass extrapolations of
finite-volume effects in QCD~\cite{fb:roma2c} and its conjunction with 
HQET~\cite{lat06:damiano}, have been proven to be feasible in quenched QCD 
and yield consistent results there, but it may turn out to be difficult to 
extended them to the dynamical case.

With $\Fbsstat=297(14)\,\MeV$ at $a\approx 0.08\,\Fm$ we find a value for
the static decay constant that is significantly larger than the $\nf=0$ 
estimate $\Fbs=194(6)\,\MeV$ quoted as the best quenched result in the 
recent review by Onogi~\cite{lat06:bphys}.
This signals a visible effect of the dynamical fermions in the two-flavour 
theory --- qualitatively in line with the outcome of other unquenched 
calculations~\cite{lat06:bphys} --- which in the end could reflect in an 
increase of $\Fbs$ and/or its $1/M$--corrections to the static limit.
Of course, these issues deserve further investigations and can only be 
settled if the two remaining sources of systematic errors are overcome, 
namely the yet slightly too large value of the sea quark mass and the lack 
of a continuum limit for $\Fbsstat$.
Therefore, the reduction of the sea quark mass as well as a determination 
of the bare static-light decay constant for a smaller lattice resolution 
are part of an ongoing project~\cite{fbstat:Nf2}.
Rather in the long term, also an active strange quark ($\nf=2+1$) will 
have to be included.
\acknowledgments
We are grateful to R.~Sommer for useful discussions and a critical reading 
of the manuscript.
We would like to thank I.~Wetzorke, R.~Hoffmann and N.~Garron for their 
help with the handling of the $\Nf=2$ configuration database and for
providing us with a basic version of the code for the computation of 
the Schr\"odinger functional correlation functions in large volume, 
respectively. 
We are also grateful to D.~Guazzini for discussions on the fit procedure
and to F.~Knechtli for communicating a few numerical details 
of~\cite{msbar:Nf2}.
This work is part of the ALPHA Collaboration research programme.
We thank NIC/DESY Zeuthen for allocating computer time on the APEmille 
computers to this project as well as the staff of the computer center
at Zeuthen for their support.
M.D.M. is supported in part by the Deutsche Forschungsgemeinschaft in the 
SFB/TR 09-03, ``Computational Particle Physics''.

\begin{appendix}
\section{Lattice actions for the static quark sector}
\label{App_lattice}
We use three discretizations of the action for static quarks, introduced 
in~\Refs{fbstat:pap1,HQET:statprec} to yield an exponential improvement of
the signal-to-noise ratio in static-light correlation functions compared to
the standard Eichten-Hill action \cite{stat:eichhill1},
\be
S_{\rm h}[U,\heavyb,\heavy]=
a^4\sum_x\heavyb(x)D_0\heavy(x) \;,
\label{action_eh}
\ee
$D_0$ being the time component of the backward lattice derivative acting on
the heavy quark field $\heavy(x)$.
These new static quark actions rely on changes of the form 
$U(x,0)\rightarrow W(x,0)$ of the parallel transporters $U(x,\mu)$ in the 
covariant derivative,
\be
D_0\heavy(x)=
\frac{1}{a}\left[\,
\heavy(x)-W^\dagger(x-a\hat{0},0)\heavy(x-a\hat{0})\,\right] \;,
\ee 
where now $W(x,0)$ is a function of the gauge fields in the immediate 
neighborhood of $x$ and $x+a\hat{0}$.
In the numerical simulations and the data analysis underlying the present
work we have considered, among the possible choices\footnote{
The sensible choice of the parallel transporters is guided by the demand of
preserving on the lattice those symmetries of the static theory, which
guarantee that the universality class and the $\Or(a)$ improvement are 
unchanged w.r.t.~the Eichten-Hill action \cite{fbstat:pap1,HQET:statprec}.
} 
for $W(x,0)$, the regularized actions 
\bea
W(x,0)=V(x,0)\left[\,
\frac{g_0^2}{5}+\left(\third\,\Tr\,V^\dagger(x,0)V(x,0)\right)^{1/2}
\,\right]^{-1}
& \quad\Rightarrow\quad &
S^\text{s}\lh \;, 
\label{action_sofx} \\
W(x,0)=V_{\rm HYP}(x,0) 
& \quad\Rightarrow\quad & 
S^\text{HYP}\lh \;,
\label{action_hyp}
\eea
where $V(x,0)$ is the average of the six staples around the link $U(x,0)$
and $V_{\rm HYP}(x,0)$ the HYP-link, the latter being a function of the gauge 
links located within a hypercube~\cite{HYP:HK01}.
The `HYP-smearing' involved in the construction of the HYP-link requires to
further specify a triplet of coefficients, the two choices
$(\alpha_1,\alpha_2,\alpha_3)=(0.75,0.6,0.3)$ and
$(\alpha_1,\alpha_2,\alpha_3)=(1.0,1.0,0.5)$ of which were motivated in
\cite{HYP:HK01} and \cite{HQET:statprec}, respectively, and define the
associated static actions $S^\text{HYP1}\lh$ and $S^\text{HYP2}\lh$.
The discretization $S^\text{s}\lh$ is inspired by the SU(3) one-link 
integral, for which $W(x,0)$ in the form of \eq{action_sofx} is an
approximation.
For more details the reader may consult~\Ref{HQET:statprec}.
In the text, we will also frequently distinguish these three static actions
by just referring to them as `s', HYP1 and HYP2. 

While the largest improvement in statistical precision is actually achieved 
for the action HYP2, it is observed \cite{fbstat:pap1,HQET:statprec} that
generally with all the proposed discretizations at least an order of
magnitude in the signal-to-noise ratios of B-meson correlation functions in 
the static approximation can be gained at time separations around 
$x_0\approx 1.5\,\Fm$ w.r.t.~the action (\ref{action_eh}) and that, in 
addition, the statistical errors grow only slowly as $x_0$ is increased.
Even more importantly, quite the same small scaling violations in the 
$\Or(a)$ improved theory are encountered with the new 
discretizations \cite{HQET:statprec}.

In that reference, also the (regularization dependent) improvement 
coefficient $\castat$ multiplying the $\Or(a)$ correction (\ref{dastat}) to 
the static-light axial current has been numerically determined in one-loop 
order of perturbation theory, and we here reproduce the corresponding 
expansions for the three static quark discretizations at our disposal:
\be
\castat(g_0)=\ca^{\rm stat,(1)}\times g_0^2
\quad \mbox{with} \quad
\ca^{\rm stat,(1)}=
\left\{\begin{array}{lll}
0.0072(4) && \mbox{for $S^\text{s}\lh$} \\
0.039(4)  && \mbox{for $S^\text{HYP1}\lh$} \\
0.220(14) && \mbox{for $S^\text{HYP2}\lh$} 
\end{array}\right. \;.
\label{ca_stat_new}
\ee

\section{Simulation results for $\zastat$}
\label{App_results}
In tables~\ref{tab:zastat1_res} and \ref{tab:zastat2_res} we collect the 
bare parameters and results of our simulations to compute $\zastat$.
The pairs $(L/a,\beta)$ and the associated values of the critical hopping
parameter, $\kappa=\hopc$, were already known from the non-perturbative
computation of the running of the Schr\"odinger functional coupling 
itself \cite{alpha:Nf2_2}.

In order to extract the lattice step scaling function $\SigmaAstat$ 
according to \eq{SSF_lat}, simulations on lattices with linear extensions 
$L/a$ and $2L/a$ are required.
At the three lowest couplings $\gbar^2(L)$, the runs have been performed 
using the one-loop value of the boundary $\Oa$ improvement coefficient 
$\ct$ in the gauge sector \cite{alpha:SU3}, 
\be
\ct^{\rm 1-lp}(g_0)=1-0.051\,g_0^2 \;,
\ee
except for $\beta=7.542$ at $L/a=6$ and $\beta=7.7206$ at $L/a=8$. 
For these parameters as well as for the larger couplings the two-loop value 
of $\ct$ \cite{impr:ct_2loop},
\be
\ct^{\rm 2-lp}(g_0)=1-0.051\,g_0^2-0.030\,g_0^4 \;,
\ee
has been employed.
At the third lowest coupling, $u\approx1.5$, we checked at $L/a=6$ that 
there is no significant difference in $\SigmaAstat$ using the one- or 
two-loop value for $\ct$.
This is also illustrated by the additional data points included 
in \fig{fig:Sigma_zastat}.
For decreasing $a/L$, the effect of the accuracy in $\ct$ on the results is
expected to become even smaller.\footnote{
The SF-specific boundary $\Oa$ improvement coefficient that involves 
the quark fields, $\cttil$, was set to its one-loop value \cite{impr:pap2}
throughout.
}
The contribution to the error of $\SigmaAstat$ induced by the uncertainty 
in the coupling $u$ (which can be estimated with the aid of the one-loop
result $\ln(2)\gamma_0$ for the derivative of $\sigmaAstat$ with respect 
to $u$) is negligible compared to the statistical error of $\SigmaAstat$.

\Tabs{tab:sigastatsepfit1_res} and \ref{tab:sigastatsepfit2_res}
list the results of the alternative continuum limit extrapolations of the 
lattice step scaling function mentioned within the discussion 
in \Sect{Sec_running_CL}, which were performed as separate fits of the
three data sets corresponding to the static actions s, HYP1 and HYP2.

Finally, we summarize in~\tab{tab:zastatpoly_res} the coefficients 
$z_i,f_i$ of the polynomial interpolations as functions 
of $5.2\leq\beta=6/g_0^2\leq 5.4$, 
\bea
\zastat(g_0,\lmax/a)=\sum_{i=0}^2z_i\,(\beta-5.2)^i \;,\quad
\ZPhi(g_0)=\sum_{i=0}^2f_i\,(\beta-5.2)^i \;,
\label{poly_zastat}
\eea
of the numerical results on the renormalization 
factors $\zastat(g_0,\lmax/a)$ and $\ZPhi(g_0)$ tabulated
in~\sect{Sec_match}.
The statistical uncertainty to be taken into account when using these
formulae varies between 0.1\% ($\beta=5.2$) and about 0.3\% ($\beta=5.4$).
Only in the case of $\ZPhi$, the additional 0.8\% error of its 
regularization independent part (\ref{Phiratio_res}) needs to be included.

\vspace{1.0cm}
%
\TABLE{
\centering
\renewcommand{\arraystretch}{1.25}
\begin{tabular}{lrrrlll}                                                        \toprule
 & & & & \multicolumn{3}{c}{$S^\text{s}\lh$} \\ \cmidrule(lr{.75em}){5-7}
 $\gbsq(L)$ & \clmc{$\beta$} & \clmc{$\kappa$} & $L/a$ & \clmc{$\zastat(L/a)$} & \clmc{$\zastat(2L/a)$} & \multicolumn{1}{c}{$\SigmaAstat(u,a/L)$}  \\ \midrule
 0.9793  & 9.50000  & 0.131532  &  6{ } & 0.9396(5)  & 0.9190(7)  & 0.9782(8)  \\
         & 9.73410  & 0.131305  &  8{ } & 0.9321(6)  & 0.9146(9)  & 0.9813(11) \\
         &10.05755  & 0.131069  & 12{ } & 0.9253(4)  & 0.9083(7)  & 0.9816(9)  \\ \midrule
 1.1814  & 8.50000  & 0.132509  &  6{ } & 0.9277(4)  & 0.9030(9)  & 0.9733(10) \\
         & 8.72230  & 0.132291  &  8{ } & 0.9195(7)  & 0.8997(10) & 0.9785(13) \\
         & 8.99366  & 0.131975  & 12{ } & 0.9115(4)  & 0.8903(11) & 0.9768(12) \\ \midrule
 1.5031  & 7.50000  & 0.133815  &  6{ } & 0.9095(6)  & 0.8783(14) & 0.9658(17) \\
         & 8.02599  & 0.133063  & 12{ } & 0.8922(8)  & 0.8649(14) & 0.9694(19) \\ \cmidrule(lr){2-7}
 1.5078  & 7.54200  & 0.133705  &  6{ } & 0.9110(6)  & 0.8780(9)  & 0.9638(12) \\
         & 7.72060  & 0.133497  &  8{ } & 0.9012(11) & 0.8736(13) & 0.9694(18) \\ \midrule
 2.0142  & 6.60850  & 0.135260  &  6{ } & 0.8843(9)  & 0.8422(11) & 0.9525(15) \\
         & 6.82170  & 0.134891  &  8{ } & 0.8752(15) & 0.8343(16) & 0.9533(24) \\
         & 7.09300  & 0.134432  & 12{ } & 0.8636(11) & 0.8219(17) & 0.9518(23) \\ \midrule
 2.4792  & 6.13300  & 0.136110  &  6{ } & 0.8623(11) & 0.8021(18) & 0.9302(24) \\
         & 6.32290  & 0.135767  &  8{ } & 0.8506(16) & 0.7984(25) & 0.9386(34) \\
         & 6.63164  & 0.135227  & 12{ } & 0.8417(9)  & 0.7894(24) & 0.9379(32) \\ \midrule
 3.3340  & 5.62150  & 0.136665  &  6{ } & 0.8310(13) & 0.7450(22) & 0.8965(29) \\
         & 5.80970  & 0.136608  &  8{ } & 0.8152(14) & 0.7367(36) & 0.9037(47) \\
         & 6.11816  & 0.136139  & 12{ } & 0.8061(14) & 0.7333(35) & 0.9097(46) \\ \bottomrule
\end{tabular}
\caption{
Results for the step scaling function $\SigmaAstat$ with discretization 
$S^\text{s}\lh$.
}\label{tab:zastat1_res}
}
%
\begin{landscape}
\TABLE{
\centering
\renewcommand{\arraystretch}{1.25}
\begin{tabular}{lrrrllllll}                                                     \toprule
 & & & & \multicolumn{3}{c}{$S^\text{HYP1}\lh$} & \multicolumn{3}{c}{$S^\text{HYP2}\lh$} \\ \cmidrule(lr{.75em}){5-7}\cmidrule(lr{.75em}){8-10}
 $\bar{g}^2(L)$ & \clmc{$\beta$} & \clmc{$\kappa$} & \clmc{$L/a$} &  
 \clmc{$\zastat(L/a)$} & \clmc{$\zastat(2L/a)$} & \clmc{$\SigmaAstat(u,a/L)$} & 
 \clmc{$\zastat(L/a)$} & \clmc{$\zastat(2L/a)$} & \clmc{$\SigmaAstat(u,a/L)$}\\ \midrule
 0.9793  & 9.50000  & 0.131532  &  6{ } & 0.9360(5)  & 0.9167(6)  & 0.9794(8)  & 0.9503(5)  & 0.9334(6)  & 0.9822(7)  \\
         & 9.73410  & 0.131305  &  8{ } & 0.9293(5)  & 0.9124(9)  & 0.9818(11) & 0.9446(5)  & 0.9291(8)  & 0.9837(10) \\
         &10.05755  & 0.131069  & 12{ } & 0.9229(3)  & 0.9064(7)  & 0.9821(9)  & 0.9387(3)  & 0.9228(7)  & 0.9831(9)  \\ \midrule
 1.1814  & 8.50000  & 0.132509  &  6{ } & 0.9242(4)  & 0.9009(8)  & 0.9748(10) & 0.9407(3)  & 0.9202(8)  & 0.9782(9)  \\
         & 8.72230  & 0.132291  &  8{ } & 0.9167(6)  & 0.8976(10) & 0.9792(12) & 0.9342(6)  & 0.9162(10) & 0.9807(12) \\
         & 8.99366  & 0.131975  & 12{ } & 0.9092(4)  & 0.8883(10) & 0.9770(12) & 0.9272(4)  & 0.9068(10) & 0.9780(11) \\ \midrule
 1.5031  & 7.50000  & 0.133815  &  6{ } & 0.9065(5)  & 0.8771(13) & 0.9675(16) & 0.9265(5)  & 0.8998(12) & 0.9712(14) \\
         & 8.02599  & 0.133063  & 12{ } & 0.8899(7)  & 0.8634(14) & 0.9703(18) & 0.9108(7)  & 0.8845(13) & 0.9712(18) \\ \cmidrule(lr){2-10}
 1.5078  & 7.54200  & 0.133705  &  6{ } & 0.9084(5)  & 0.8771(9)  & 0.9655(11) & 0.9283(5)  & 0.8998(8)  & 0.9692(10) \\
         & 7.72060  & 0.133497  &  8{ } & 0.8991(10) & 0.8726(13) & 0.9706(17) & 0.9201(9)  & 0.8949(12) & 0.9726(16) \\ \midrule
 2.0142  & 6.60850  & 0.135260  &  6{ } & 0.8824(8)  & 0.8419(11) & 0.9541(14) & 0.9079(7)  & 0.8700(10) & 0.9583(13) \\
         & 6.82170  & 0.134891  &  8{ } & 0.8740(14) & 0.8341(16) & 0.9544(23) & 0.8997(13) & 0.8610(16) & 0.9570(21) \\
         & 7.09300  & 0.134432  & 12{ } & 0.8624(10) & 0.8218(17) & 0.9529(22) & 0.8874(9)  & 0.8467(16) & 0.9541(20) \\ \midrule
 2.4792  & 6.13300  & 0.136110  &  6{ } & 0.8619(10) & 0.8045(16) & 0.9333(23) & 0.8924(9)  & 0.8366(16) & 0.9375(21) \\
         & 6.32290  & 0.135767  &  8{ } & 0.8505(15) & 0.7998(24) & 0.9405(33) & 0.8807(14) & 0.8304(24) & 0.9429(32) \\
         & 6.63164  & 0.135227  & 12{ } & 0.8417(8)  & 0.7905(23) & 0.9392(31) & 0.8694(8)  & 0.8178(22) & 0.9406(29) \\ \midrule
 3.3340  & 5.62150  & 0.136665  &  6{ } & 0.8324(12) & 0.7497(20) & 0.9007(28) & 0.8707(11) & 0.7896(20) & 0.9069(27) \\
         & 5.80970  & 0.136608  &  8{ } & 0.8177(13) & 0.7390(34) & 0.9037(44) & 0.8542(13) & 0.7730(35) & 0.9049(43) \\
         & 6.11816  & 0.136139  & 12{ } & 0.8073(13) & 0.7334(33) & 0.9085(44) & 0.8398(13) & 0.7630(32) & 0.9086(42) \\ \bottomrule
\end{tabular}
\caption{
Results for the step scaling function $\SigmaAstat$ with discretizations 
HYP1 and HYP2.
}\label{tab:zastat2_res}
}
\end{landscape}
%
\TABLE[t]{
\centering
\renewcommand{\arraystretch}{1.25}
\begin{tabular}{cclclcl} \toprule
 $u$ & & \clmc{$\sigma^{\text{stat}}_{A,\text{s}}(u)$}
     & & \clmc{$\sigma^{\text{stat}}_{A,\text{HYP1}}(u)$}
     & & \clmc{$\sigma^{\text{stat}}_{A,\text{HYP2}}(u)$} \\ \midrule
  0.9793 & & 0.9832(13) & & 0.9834(13) & & 0.9838(12) \\
  1.1814 & & 0.9793(16) & & 0.9791(16) & & 0.9790(15) \\
  1.5031 & & 0.9706(26) & & 0.9712(25) & & 0.9712(24) \\
  2.0142 & & 0.9522(25) & & 0.9530(24) & & 0.9533(23) \\
  2.4792 & & 0.9423(36) & & 0.9428(35) & & 0.9433(34) \\
  3.3340 & & 0.9138(57) & & 0.9103(55) & & 0.9076(52) \\ \bottomrule
\end{tabular}
\caption{
Results of the continuum limit extrapolation of the lattice step scaling 
function $\SigmaAstat(a/L,u)$ to $\sigmaAstat(u)$, fitting the data from 
our three discretizations separately at all available lattice resolutions 
as a function linear in $(a/L)^2$.
}\label{tab:sigastatsepfit1_res}
}
%
\TABLE[b]{
\centering
\renewcommand{\arraystretch}{1.25}
\begin{tabular}{cclclcl} \toprule
 $u$ & & \clmc{$\sigma^{\text{stat}}_{A,\text{s}}(u)$}
     & & \clmc{$\sigma^{\text{stat}}_{A,\text{HYP1}}(u)$}
     & & \clmc{$\sigma^{\text{stat}}_{A,\text{HYP2}}(u)$} \\ \midrule
  0.9793 & & 0.9815(8)  & & 0.9820(7)  & & 0.9833(7)  \\
  1.1814 & & 0.9776(9)  & & 0.9781(9)  & & 0.9793(8)  \\
  1.5031 & & 0.9694(13) & & 0.9704(12) & & 0.9720(12) \\
  2.0142 & & 0.9525(15) & & 0.9536(14) & & 0.9555(14) \\
  2.4792 & & 0.9383(21) & & 0.9398(21) & & 0.9417(20) \\
  3.3340 & & 0.9067(34) & & 0.9061(33) & & 0.9068(32) \\ \bottomrule
\end{tabular}
\caption{
As in \tab{tab:sigastatsepfit1_res} but upon omitting the $L/a=6$ 
data and fitting to a constant.
}\label{tab:sigastatsepfit2_res}
}
%
\TABLE[t]{
\centering
\renewcommand{\arraystretch}{1.0}
\def\onelp{\small{1-lp}}
\begin{tabular}{ccccr@{.}lcr@{.}lcl} \toprule
 $\Sstat$ & & $i$ & & \multicolumn{2}{c}{$z_i$} & & 
 \multicolumn{2}{c}{$f_i$} & & $\castat$ \\ \midrule
 s    & & 0 & &  $ 0$&$7920$  & &  $ 0$&$6970$  & &  \onelp  \\
      & & 1 & &  $-0$&$0393$  & &  $-0$&$0358$  & &          \\
      & & 2 & &  $-0$&$1434$  & &  $-0$&$1212$  & &          \\[0.75ex]
      & & 0 & &  $ 0$&$7941$  & &  $ 0$&$6988$  & &  0       \\
      & & 1 & &  $-0$&$0421$  & &  $-0$&$0370$  & &          \\
      & & 2 & &  $-0$&$1369$  & &  $-0$&$1202$  & &          \\\midrule
 HYP1 & & 0 & &  $ 0$&$7962$  & &  $ 0$&$7007$  & &  \onelp  \\
      & & 1 & &  $-0$&$0276$  & &  $-0$&$0257$  & &          \\
      & & 2 & &  $-0$&$1869$  & &  $-0$&$1591$  & &          \\[0.75ex]
      & & 0 & &  $ 0$&$8073$  & &  $ 0$&$7104$  & &  0       \\
      & & 1 & &  $-0$&$0364$  & &  $-0$&$0313$  & &          \\
      & & 2 & &  $-0$&$1753$  & &  $-0$&$1586$  & &          \\\midrule
 HYP2 & & 0 & &  $ 0$&$8446$  & &  $ 0$&$7432$  & &  \onelp  \\
      & & 1 & &  $-0$&$0448$  & &  $-0$&$0393$  & &          \\
      & & 2 & &  $-0$&$1934$  & &  $-0$&$1687$  & &          \\[0.75ex]
      & & 0 & &  $ 0$&$9000$  & &  $ 0$&$7920$  & &  0       \\
      & & 1 & &  $-0$&$0691$  & &  $-0$&$0604$  & &          \\
      & & 2 & &  $-0$&$2318$  & &  $-0$&$2056$  & &          \\ \bottomrule
\end{tabular}
\caption{
Coefficients of the interpolating polynomials of the renormalization
factors in~\eq{poly_zastat}. 
Uncertainties are discussed in the text.
}\label{tab:zastatpoly_res}
}
%
\clearpage
\end{appendix}
\bibliography{lattice_ALPHA}
\bibliographystyle{JHEP-2}
\end{document}